\documentclass[twocolumn,nofootinbib,aps,floatfix,superscriptaddress,preprintnumbers]{revtex4}

\usepackage{graphicx}

\include{myBabarsym}
\newcommand{\bei}{\begin{itemize}}
\newcommand{\eei}{\end{itemize}}
\newcommand{\beq}{\begin{equation}}
\newcommand{\eeq}{\end{equation}}
\newcommand{\beqn}{\begin{eqnarray}}
\newcommand{\eeqn}{\end{eqnarray}}
\newcommand{\beqns}{\begin{eqnarray*}}
\newcommand{\eeqns}{\end{eqnarray*}}

\newcommand{\intl}{\int\limits}

\newcommand{\hmmm}{\hspace{-0.3cm}}
\newcommand\PRL{{ Phys. Rev. Lett.}}

\newcommand\ea{{\em et al.}\xspace}

\newcommand\PRD{{ Phys. Rev.}}
\newcommand\Amptpbar{\kern 0.18em\overline{\kern -0.18em {\cal A}}_{3\pi}}

\newcommand\Amptpbarkappa{\kern 0.18em\overline{\kern -0.18em A}^{\kappa}{}}
\newcommand\Amptpbarsigma{\kern 0.18em\overline{\kern -0.18em A}^{\sigma}{}}

\renewcommand\Re{{\rm Re}}
\renewcommand\Im{{\rm Im}}
\newcommand\CL{{\rm CL}}

\newcommand\Nbpm{{\kern 0.18em\overline{\kern -0.18em N}}^{+-}}
\newcommand\Nbmp{{\kern 0.18em\overline{\kern -0.18em N}}^{-+}}
\newcommand\pdf{PDF\xspace}
\newcommand\pdfs{PDFs\xspace}

\newcommand\MA{MA\xspace}
\newcommand\RI{RI\xspace}
\newcommand\ES{ES\xspace}
\newcommand\RIt{\mathrm{RI}_\tau\xspace}
\newcommand\alphaeff{\alpha_{\rm eff}}

\newcommand\Cpizpiz{{C_{\pi\pi}^{00}}}

\newcommand\Cpipi{{C_{\pi\pi}^{+-}}}
\newcommand\Spipi{{S_{\pi\pi}^{+-}}}

\newcommand\BRpipi{{\cal B}^{+-}_{\pi\pi}}
\newcommand\BRpippiz{{\cal B}^{+0}_{\pi\pi}}
\newcommand\BRpizpiz{{\cal B}^{00}_{\pi\pi}}

\newcommand\BRpmb{{\cal \kern 0.18em\overline{\kern -0.18em  B}}{}_{\rho\pi}^{+-}}
\newcommand\BRmpb{{\cal \kern 0.18em\overline{\kern -0.18em  B}}{}_{\rho\pi}^{-+}}

\newcommand\BRipmb{{\cal \kern 0.18em\overline{\kern -0.18em  B}}{}_{\rho^+\pi^-}}
\newcommand\BRimpb{{\cal \kern 0.18em\overline{\kern -0.18em  B}}{}_{\rho^-\pi^+}}

\newcommand\Abar{\kern 0.18em\overline{\kern -0.18em A}{}}

\newcommand\ie{{i.e.}\xspace} 
 
\newcommand\eg{{e.g.}\xspace}

\newcommand\Lik{{\cal L}}

\newcommand\Ndof{N_{\rm dof}}

\newcommand\ProbCERN{{\rm Prob}}

\newcommand\MNP{\ensuremath{M_{\rm NP}}\xspace}
\newcommand\MNPb{\kern 0.18em\overline{\kern -0.18em M}_{\rm NP}{}\xspace}

\begin{document}


\preprint{
\vbox{
\hbox{CERN-OPEN-2006-029}
\hbox{CPT-P56-2006}
\hbox{LAL 06-122}
\hbox{LAPP-2006-02}
}}

\vspace*{1mm}

\title{\boldmath Bayesian Statistics at Work: the Troublesome Extraction of the CKM Phase $\alpha$}

\author{J.~Charles}
\affiliation{CPT, Luminy Case 907, F-13288 Marseille Cedex 9, France}

\author{A.~H\"ocker}
\affiliation{CERN, CH-1211 Geneva 23, Switzerland}

\author{H.~Lacker}
\affiliation{TU Dresden, IKTP, D-01062 Dresden, Germany}

\author{F.R. Le Diberder}
\affiliation{LAL, CNRS/IN2P3, Universit\'e Paris-Sud 11, B\^at. 200, BP 34, F-91898 Orsay Cedex, France}

\author{S. T'Jampens}
\affiliation{LAPP, CNRS/IN2P3, Universit\'e de Savoie, 9 Chemin de Bellevue, BP 110, F-74941 Annecy-le-Vieux Cedex, France}

\date{\today}

\begin{abstract}
  In Bayesian statistics, one's prior beliefs about underlying model parameters 
  are revised with the information content of observed data 
  from which, using Bayes' rule, a posterior belief is obtained.   
  A non-trivial example taken from the isospin analysis of $B \to PP$ ($P =
  \pi$ or $\rho$) decays in heavy-flavor physics is chosen to illustrate the
  effect of the naive ``objective'' choice of flat priors in a
  multi-dimensional parameter space in presence of mirror solutions. It is
  demonstrated that the posterior distribution for the parameter of interest,
  the phase $\alpha$, strongly depends on the choice of the parameterization
  in which the priors are uniform, and on the validity range in which the
  (un-normalizable) priors are truncated. We prove that the most probable values found by the 
  Bayesian treatment do not coincide with the explicit analytical solutions, in 
  contrast to the frequentist approach. It is also shown in the appendix that 
  the $\alpha \to 0$ limit cannot be consistently treated in the Bayesian paradigm, 
  because the latter violates the physical symmetries of the problem.
\end{abstract}

\maketitle

%
%

\section{Introduction}

In Bayesian statistics, probability is a measure of one
person's state of knowledge (also called degree of belief) of the plausibility
of a proposition given incomplete knowledge at a given time. Another person
may have a different degree of belief in the same proposition, and so have a
different probability. The only constraint is that the probabilities chosen
by  a single person should be consistent (``coherent''): they should obey all
the axioms of probability~{\cite{bayesbook}}. Bayes' rule is understood as a
revision process, by which a prior probability is changed into a new one, the
posterior probability, due to input information provided by the data. 

The result of any inference problem is the posterior distribution of the quantity
of interest. Bayesian models require the specification of prior distributions
for all unknown parameters, expressing the actual personal degrees of belief
based on all the available information prior to updating one's degree of
belief with the information content of the data. 
In the case where prior knowledge about model parameters 
is unavailable, the specification of prior distributions is never 
unequivocal.

Neither Bayesian statistics nor any other framework provides fundamental rules
for obtaining the prior probability about the parameters.\footnote
{
  It is not surprising that no rules are given because knowledge is a very poorly defined
  concept. In the personalistic Bayesian approach 
  developed by F.P. Ramsey, B. de Finetti and L.J. Savage, personal degrees of
  belief are represented numerically by betting quotients: one should assign and
  manipulate probabilities so that one cannot be made a sure loser in betting
  based on them~\cite{probaphilo1}. Betting cannot be used to measure the strength
  of someone's belief in a universal scientific law or theory~\cite{Gillies88}.
} 
The specification of prior distribution
may be possible in some simple cases but is impractical in complicated
problems if there are many parameters.
In practice, especially when nothing or very little is known about the
parameters, most Bayesian analyses are performed with so-called
``non-informative priors''~{\cite{Kass}}. An obvious candidate for a
non-informative prior is to use a flat prior. The notion of a flat prior is
not well-defined because a flat prior of one parameter does not imply a flat
prior on a transformed version of that parameter. Prior density distributions
are not transformation-invariant, because they depend on the metric.
For example, a uniform distribution of $\theta$ does not lead to a uniform
distribution of $\mu = \sin (\theta)$. Thus, there is a fair amount of
arbitrariness in how ignorance is parameterized, which will affect the
posterior probability and hence the result. Moreover, there is a fundamental
difference rarely acknowledged between knowing that a uniform prior
probability distribution in the range $[\theta_a, \theta_b]$ has been assigned
to the value of a parameter $\theta$ as a result of positive knowledge, and
not knowing anything about $\theta$ with the exception of its
admissible range. These are two fundamentally different states of knowledge. 
It is often claimed, though without any proof, that the relative prior dependence
of the posterior probability distribution is reduced as the statistical
information from the measured data is increased.

In the physical sciences, the invariance of conclusions drawn from data under
a particular parameter choice is a fundamental concept. Furthermore, it is
questionable~{\cite{Mayo}} ``whether scientists have prior degrees of beliefs
in the hypotheses they investigate and whether, even if they do, it is
desirable to have them figure centrally in learning from data to science. In
science, it seems, we want to know what the data are saying, quite apart from
the opinions we start out with.''

The mathematical content of this paper being rather simple, its purpose is to illustrate with
a concrete use case how strongly prior-dependent the Bayesian treatment can be. The
chosen example is taken from the field of particle physics, more specifically
from recent results discussed in the domain of  \CP violation. The extraction of the
CKM phase $\alpha$ from the isospin analysis of $B \to PP$ ($P = \pi$ or
$\rho$) decays is used as an illustration of a Bayesian analysis at work with
flat priors in a multi-dimensional parameter space in presence of mirror
solutions. Troublesome results are obtained~{\cite{UTfit1}}.

We begin by introducing the analysis formalism and the statistical approaches used 
to interpret the experimental results. 
We present the results of the so-called $\B\to\pi\pi$ {\em isospin analysis} 
(see below) in several parameterizations
finding that the Bayesian method leads to very different conclusions depending
on the choice of the parameterization. We then explicit a simpler
two-dimensional example that bears some similarities with the extraction of
$\alpha$, namely the existence of mirror solutions, and show why the Bayesian
treatment amounts to an unacceptable interpretation of fundamental physics
parameters. Finally in appendix we explain in detail the role of the $\alpha
\to 0$ limit, its associated mathematical properties and its relation to 
\CP violation and new physics. We show that the Bayesian treatment leads to an unrecoverable
divergence when one parameterizes the Standard Model amplitudes by their real and imaginary parts,
\ie, when one uses the parameterization that is the most natural from the point of view of the 
computation of  Feynman diagrams.
Readers well-aware of the basics of the frequentist and Bayesian statistical treatments may skip 
Sections~\ref{sec:freq}, \ref{sec:bayes} and \ref{sec:prior}.

%
%

\section{Analysis Formalism}\label{sec:formalism}

The experimental framework consists of the measurement of six observables:
three branching fractions and three asymmetries (see, \eg, Ref.~\cite{babarphysbook}).
The three branching fractions are related to the three \B meson decays:
$B^{0}\to\pi^+\pi^-$, $B^{0}\to\pi^0\pi^0$, $B^\pm\to\pi^\pm\pi^0$,
where the average is implicitly taken between the two \CP-conjugate decays;
$B^0$ and $\Bzb$, and $B^+$ and $B^-$, respectively.
The three asymmetries are quantities which would vanish in the absence
of \CP violation (charge conjugation times spatial parity):
theses quantities modulate the time dependence of the neutral \B-meson decays. 
They are denoted $S^{+-}$, $C^{+-}$ and $C^{00}$.

The use of the isospin analysis to extract fundamental parameters from the 
$B\to\pi\pi$ observables is a well known problem that was first solved in 
1990~\cite{GL}. Assuming isospin symmetry (an approximate SU(2) flavor symmetry 
of the strong interaction, which is known to hold to better than a few percent 
accuracy), the two pion final state can be represented as a superposition of 
isospin zero ($I=0$) and isospin two ($I=2$) eigenstates. Within the
Standard Model, the $B^0$-decay amplitudes can then be written as 
\begin{eqnarray}\label{SMparam}
A^{+-} & \equiv & A(B^0\to\pi^+\pi^-) = e^{-i\alpha}T^{+-}+P \,, \nonumber\\
\sqrt{2}A^{00} & \equiv & \sqrt{2}A(B^0\to\pi^0\pi^0) =
e^{-i\alpha}T^{00}-P\, ,\\
\sqrt{2}A^{+0} & \equiv & \sqrt{2}A(B^+\to\pi^+\pi^0) =
e^{-i\alpha}(T^{00}+T^{+-}) \,,\nonumber
\end{eqnarray}
where a peculiar phase convention has been taken (in which $q/p$, the $\Bz\Bzb$ 
mixing phase shift, is equal to one). The triangular relation
$\sqrt{2}A^{+0}=\sqrt{2}A^{00}+A^{+-}$ that follows
from the parameterization above is a consequence of the isospin
symmetry, and the only information that we have on the amplitudes without
any additional hypothesis; the
``$P$'' (the so-called ``penguin'') term in the neutral mode comes from $\Delta I=1/2$ operators that cannot
contribute to the $\Delta I=3/2$ charged transition, hence the absence of
a ``$P$'' term in $A^{+0}$. This notation makes explicit the presence
of \CP violation through the \CP-odd weak-interaction phase $\alpha$ (which is the
parameter of main interest), while the other (hadronic) parameters are \CP-conserving 
complex numbers. The \CP-conjugated amplitudes are thus obtained from Eq.~(\ref{SMparam}) 
by a sign flip $\alpha\to -\alpha$. In the following, the parameterization from Eq.~(\ref{SMparam}) 
will be referred to as the ``Standard Model'' parameterization.

The observables that are currently measured by the $B$-factory experiments \babar and Belle
are the \CP-averaged branching fractions $\BR^{ij}\propto (\tau_B/2)(|A^{ij}|^2+|
\Abar^{ij}|^2)/2$, where $\tau_B$ is the \Bz- or \Bp-meson lifetime depending on the modes, 
the direct \CP asymmetries $C^{ij}=(|A^{ij}|^2-|
\Abar^{ij}|^2)/(|A^{ij}|^2+|\Abar^{ij}|^2)$ and the $\Bz\Bzb$-mixing-induced \CP asymmetry
$S^{+-}=\Im(\Abar^{+-}/A^{+-})$. This adds up to six independent constraints on the six
independent parameters, namely $\alpha$ and either the modulus and argument, or equivalently
the real and imaginary parts, of $T^{+-}$, $T^{00}$ and $P$ (one overall phase being
irrelevant, it can be fixed to any value without altering the observables). 
Thus the system of equations is just constrained. It can be inverted 
explicitly~\cite{JC}, what  will be referred to in the following as the ``explicit solution'' 
parameterization.\footnote
{\label{foot:eight}
One can  extract the angle $\alpha$, up to discrete ambiguities, provided 
electroweak penguin contributions are negligible ($P^{\rm EW}=0$). 
The explicit solutions in terms of $\alpha$ are given by~\cite{LDP,thePapII}
$$
\tan\alpha=\frac{\sin(2\alphaeff)\bar c+\cos(2\alphaeff)\bar s+s}
        {\cos(2\alphaeff)\bar c-\sin(2\alphaeff)\bar s+c}~,
$$
where all quantities on the right hand side can be expressed in term of
the observables as follows:
$$
 \sin(2\alphaeff) =  \frac{\Spipi}{\sqrt{1-\Cpipi^2}}~,        
 \cos(2\alphaeff) =  \pm\sqrt{1-\sin^2(2\alphaeff)}~,
$$
$$c            =\sqrt{\frac{\tau_{B^+}}{\tau_{B^0}}}\,\frac{\frac{\tau_{B^0}}{\tau_{B^+}}
              \BRpippiz+\BRpipi(1+\Cpipi)/2-\BRpizpiz(1+\Cpizpiz)}
              {\sqrt{2\BRpipi\BRpippiz(1+\Cpipi)}}~,
$$
$$\overline{c} = \sqrt{\frac{\tau_{B^+}}{\tau_{B^0}}}\,\frac{\frac{\tau_{B^0}}{\tau_{B^+}}
               \BRpippiz+\BRpipi(1-\Cpipi)/2-\BRpizpiz(1-\Cpizpiz)}
               {\sqrt{2\BRpipi\BRpippiz(1-\Cpipi)}}~,
$$
$$
s           = \pm\sqrt{1-c^2}~, 
\overline{s}= \pm\sqrt{1-\overline c^2}~.
$$
The eightfold ambiguity for $\alpha$ in the range $[0,\pi]$ is made explicit by the three arbitrary signs.
}
It is assumed throughout this paper that all observables are obtained 
from Gaussian measurements, and that there is no source of uncertainty other than 
the unknown tree and penguin amplitudes defined above. The fit of $\alpha$ 
or of any subset of the unknown parameters is therefore a classical, well-defined, 
statistical problem.

What makes the isospin analysis an interesting example, besides its non-linearity
due to trigonometric functions, is the presence of a high-order exact degeneracy between
mirror solutions. Indeed it can be shown explicitly (see, \eg, Ref.~\cite{JC}) that there are eight
parameter sets that give exactly the same value for the observables when one restricts
$\alpha$ to the range $[0,\pi]$ (eight other sets are found trivially through the
transformation $\alpha\to\alpha +\pi$, $T^{ij}\to - T^{ij}$ that leaves the system
invariant). Even with infinite statistics, it is not possible from a given set of
measurements to determine which of the various mirror solutions is the true one.

The same formalism can be applied (to a good approximation) to the quasi two-body 
decay $B\to\rho\rho$. However in this case, only an upper bound on the
branching fraction to $\rho^0\rho^0$ is currently available, which makes the isospin 
analysis an under-constrained system. It has been shown in the literature~\cite{GQ,JC} 
that one obtains bounds on the phase $\alpha$ in such a case.

We consider another useful parameterization of the decay amplitudes, which has been 
proposed in Ref.~\cite{LDP}. It will be referred to as the ``Pivk-LeDiberder'' 
parameterization. One introduces six parameters $\alpha$, $\alpha_\mathrm{eff}$, 
$\mu$, $a$, $\bar a$ and $\Delta$ through the definitions
\begin{eqnarray}
A^{+-} &=& \mu\, a \,, \nonumber\\
A^{00} &=& \mu\, e^{i(\Delta-\alpha)}
   \left( 1-\frac{a}{\sqrt{2}}\,e^{-i(\Delta-\alpha)}\right) \,, \nonumber\\
A^{+0} &=& \mu\, e^{i(\Delta-\alpha)} , \\
\Abar^{+-} &=& \mu\, \bar a\, e^{2i\alpha_\mathrm{eff}} \,, \nonumber\\
\Abar^{00} &=& \mu\, e^{i(\Delta+\alpha)}
   \left( 1-\frac{\bar
   a}{\sqrt{2}}\,e^{-i(\Delta+\alpha-2\alpha_\mathrm{eff})}\right) \,, \nonumber\\
\Abar^{+0} &=& \mu\, e^{i(\Delta+\alpha)}\, .\nonumber
\end{eqnarray}
The above description is the simplest one that makes explicit the
two crucial ingredients of the isospin analysis, namely the triangular
relation between the amplitudes and the fact that $2\alpha$ is the
phase difference between $\Abar^{+0}$ and $A^{+0}$. 

%
%

\section{Frequentist Analysis}\label{sec:freq}

``Frequentist statistics provides the usual tools for reporting objectively the 
outcome of an experiment without needing to incorporate prior beliefs concerning 
the parameter being measured or the theory being tested. As such they are used 
for reporting essentially all measurements and their statistical uncertainties in 
High Energy Physics''~\cite{pdg2005}. The Frequentist sees probability as the 
long-run relative frequency of occurrence. 
Hence, the frequentist analysis assumes that a population mean is real, but unknown, 
and unknowable, and can only be estimated from the available data. 

We neglect in the following the occurrence of physical boundaries 
for the true values which greatly simplifies the computation of frequentist confidence levels. 
We adopt a $\chi^2$-like notation and define
\beq\label{eq:chi2Function}
   \chi^2(\theta) \equiv  -2\ln(\Lik_{\{\mathbf{x} \}}(\theta))\,,
\eeq
where the likelihood function, $\Lik$, quantifies the 
agreement between the measured observables, $\{ \mathbf{x}\}$, and their theoretical 
counterparts, $\{\mathbf{f}(\theta)\}$. The $\theta$ parameters 
are the unknowns of the theory, \eg, for the Pivk-LeDiberder
parameterization one has $\theta = \{\alpha,\alpha_{\rm eff},\mu,a,\bar a,\Delta\}$.

Under these assumptions, and neglecting experimental correlations,
the likelihood components of $\Lik$ are independent Gaussian distribution 
functions
\beq
\label{eq:thegaussian}
\Lik_{\{x_i \}}(\theta) \propto
        {1\over\sigma_i}\,
        {\rm exp}\left[-{1\over 2}
        \left({x_i-f_i(\theta)\over\sigma_i}\right)^{\!\!2\,}\right]\,,
\eeq
each with a standard deviation given by the statistical uncertainty 
$\sigma_i$ on the measurement $i$.\footnote
{
   In practice, one has to deal with correlated measurements and 
   with additional experimental and theoretical systematic uncertainties,
   which are however irrelevant for the discussion of this paper.
} 
In this case, using incomplete $\Gamma$ functions (as
computed, \eg, with "$\ProbCERN$" the well known routine from
the CERN library), 
one can infer a confidence level (\CL) from the above $\chi^2$ value as follows
\begin{eqnarray}
\label{eq':probcern}
\CL      &=&     \ProbCERN(\chi^2(\theta),\Ndof)~, \nonumber\\
         &=&     \frac{1}{\sqrt{2^{\Ndof}}\Gamma({\Ndof}/2)}
                 \intl_{\chi^2(\theta)}^\infty 
                 \hmmm e^{-t/2}t^{\Ndof/2-1}\, dt\,.
\end{eqnarray}

%
%

\section{Bayesian Analysis}\label{sec:bayes}

Bayesian probability, also named personal probability or (more often but less appropriately) subjective probability, 
represents one's degree of belief.\footnote
{
  A belief could be well- or ill-founded, in agreement or disagreement with the facts, etc., 
  it is clear that an expression or assertion of that belief has no necessary connections to the facts.
  It is not ``objective'' in the root sense of being about or dependent upon objects in the real world, 
  but is rather subjective in the sense of being about or dependent upon the psychological subject~\cite{probaphilo2}.
}
It is thus a summary of one's own opinions about an uncertain proposition, not
something inherent to the system being studied.\footnote
{
  Bayesian personal probability reflects the scientist's confidence that a hypothesis is true (among all other
  rival hypotheses). A scientist's personal probability for a hypothesis is,
  then, more a psychological fact about the scientist than an
  observer-independent fact about a hypothesis. It is not a matter of how likely
  the truth of a hypothesis actually is, but about how likely the scientist
  thinks it to be~{\cite{Strevens,Mayo}}. In other words, in the personalistic Bayesian viewpoint,
  ``the probability of $A$ is $1 / 2$'' is not a statement about $A$, it is
  a statement about the state of mind (opinion) of the person making the assertion.
  In a system which defines probability as the individual's degree of belief in a proposition, 
  it is obvious that there can be no one answer to ``what is the probability of X?'' 
  There are as many answers as there are beliefs, and no answer is better than any other (coherent) answer,
  since the individual is theoretically free to hold any opinion whatsoever~\cite{probaphilo2}.
} 
Thus, in the personalistic Bayesian approach, hypotheses or unknowns can never be directly measured 
or statistically evaluated. A personal probability statement cannot be proved or disproved, verified or falsified.\footnote
{ 
  One characteristic of the scientific method is the formulation of testable hypothesis.
  The objectivity of scientific statements rely upon the fact that they can be submitted 
  to tests in a reliable manner and with checkable assumptions~\cite{Science}.
  So, in order to exceed the level of a mere speculation, any theory of inference
  about parameters must be exposed, \ie, must be able to make predictions that can
  be verified by experiments (or falsified, in K. Popper's version of the same idea).
  Hence, adopting a set of axioms does not guarantee a success in modeling the empirical world
  --- one needs an extra argument, such as empirical verification, to justify the 
  use of any given set of axioms. The personalistic Bayesian viewpoint claims that probability statements 
  cannot be verified (because probability does not exist in an objective sense, in de Finetti's motto: ``Probability does not exist''). 
  The important point here is that probabilities obtained in this way do not correspond 
  directly to anything objectively existing in the real world. To be tested a probability proposition needs to
  be converted to a statistical proposition, which is verifiable. Probability statements are not descriptive but
  judgmental.
}


A \textit{posterior}  probability density function (\pdf), 
$P(\mathbf{\theta}|\{\mathbf{x}\})$, for a model parameter
$\theta$ having observed the data $\{\mathbf{x}\}$ is, using the Bayes' rule:
\begin{equation}
  \P(\theta |\{ \mathbf{x} \}) = \frac{ \P(\{ \mathbf{x} \}| \theta) \pi
  (\theta)}{\int \hspace{-0.25em}  \P(\{ \mathbf{x} \}| \theta) \pi (\theta) d
  \theta} \,,
\end{equation}
where $\P (\{ \mathbf{x} \}| \theta)$ is nothing more than the likelihood
function\footnote
{
  $\P(\mathbf{x} | \theta)$ is a probability (discrete data)
  or a \pdf  (continuous data) \textit{as a function of the data} and all possible
  data are considered, including the data not observed. If the data are
  considered fixed (at the measured value), then $\P ( \mathbf{x} | \theta)$ is
  no longer a probability or a  \pdf, it becomes the \textit{likelihood function},
  $ \Lik_{\{ \mathbf{x} \}} (\theta) \equiv P (\{ \mathbf{x} \}| \theta)$, a function of the true model
  parameter $\theta$~{\cite{James03}}. The likelihood function is not the \pdf  of
  $\theta$, given $\{ \mathbf{x} \}$. To turn the likelihood function into the \pdf
  $P (\theta |\{ \mathbf{x} \})$, one needs to invoke the Bayes' rule where a
  prior \pdf is mandatory. 
} 
($\Lik_{\{ \mathbf{x} \}}(\theta)$) of the true parameter value $\theta$ taken at the observed data $\{\mathbf{x}\}$. 
The posterior \pdf in $\theta$ is obtained by multiplying $\Lik$ by the 
\textit{prior}  \pdf $\pi(\theta)$, the probability density function for the model
parameter $\theta$. In essence, the prior is reweighed according to the likelihood of the data.\footnote
{
  In Bayesian probability, one does not try to test or refute one's prior probabilities, one simply changes them 
  into posterior probabilities by Bayesian conditionalization. If the initial assumptions are seriously wrong 
  in some respects, then not only will the prior probability function be inappropriate, but all the conditional 
  probabilities generated from it in the light of new evidence will also be inappropriate. To obtain reasonable 
  probabilities in such circumstances, it will be necessary to change one's prior probability in a much more drastic 
  fashion than Bayesian probability allows, and, in effect, introduce a new prior probability function~\cite{probaphilo1}.
  The important point is that even when one's degree of belief changes with new evidence, in no way does it 
  show one's previous degree of belief to have been mistaken.
  Furthermore, no proof is required for the posterior distribution to have desirable properties. 
  The personalistic Bayesian philosophy not only fails to make such a recommendation 
  but asserts that this cannot be done at all.
}

It should be emphasized here that the probability associated with the value of a model parameter cannot be
interpreted meaningfully as a frequency of an outcome of a repeatable
experiment. Instead, it is understood to reflect the degree of belief
that the parameters have particular values. Stated otherwise, since the 
parameter $\theta$ is not a random variable,\footnote
{
	Following the Bayesian paradigm, a  probability density distribution 
	to $\theta$ is assigned to express one's uncertainty, not to attribute 
	randomness to $\theta$~\cite{ohagan}. 
} 
the probability distribution for $\theta$ is not a
probability distribution in the usual frequency sense, \ie, one cannot sample 
from this distribution and obtain various values for $\theta$~\cite{Porter}. 
However, the prior \pdf $\pi(\theta)$ can be defined
with the formal rules of probabilities and quantifies one's
degree of belief about the parameter before carrying out the experiment,
\ie, no matter what the data are. There is no fundamental recipe 
for assigning a priori probabilities to parameters. Bayes' rule, after 
choosing a certain prior $\pi(\theta)$, only states how the a posteriori 
probability changes in the light of the existing
experimental data. In other words, Bayesian posterior \pdf depends not
only on the observation itself, but also on the state of knowledge and
beliefs of the observer. As a consequence, the posterior \pdf
by itself does not in general provide a useful summary of the result
of the experiment, as it convolves the data with the personal
beliefs needed to construct the prior \pdf. 

In the case of more than one parameter, $\theta_{1}, \ldots, \theta_{m}$, 
the a posteriori \pdf of, say the parameter $\theta_1$, is obtained
by integrating out the parameters $\theta_{2}, ..., \theta_{m}$ to get
the \textit{marginal} \pdf
\begin{equation}\label{eq:marginalpdf}
\P(\theta_{1}|\{\mathbf{x}\}) =
\int\! \P(\theta_{1},\theta_{2},
\ldots, \theta_{m} | \{\mathbf{x}\}) 
 d \theta_{2}\ldots d \theta_{m}\,.
\end{equation}
By doing so, one has chosen a certain parameterization ($\theta_{1}, \ldots, \theta_{m}$) 
and a corresponding metric ($ d \theta_{1}\ldots d \theta_{m}$).  

%
%
\section{Priors}\label{sec:prior}

Non-informative prior distributions are generally improper (they do not
normalize) when the parameter space is not compact which may lead to an
improper posterior. However, a posterior must
always be proper, in other words, it must be a probability (discrete
parameter) or a \pdf (continuous parameter). To remedy this problem
what is done in practice is to truncate\footnote
{
   Use of ``vague proper prior'' in such situations will formally result 
   in proper posterior distributions, but these posteriors will essentially 
   be meaningless if the limiting improper prior had resulted in an 
   improper posterior distribution. \label{truncate}
} 
the range of the prior. However, the ranges for the prior \pdf with $\pi(\theta) \ne 0$ 
restrict the allowed range for the posterior \pdfs in $P(\theta|\{\mathbf{x}\})$. 
Hence, it has to be verified that the a priori ranges do not introduce a cut in
the posterior \pdf. If this, however, is the case one needs to either enlarge 
the ranges of the prior \pdf where $\pi(\theta) \ne 0$, or to justify the 
ranges used.

In an ideal case, the posterior \pdf should not depend on 
the prior \pdf but only on the experimental likelihood. However, in reality, the choice of prior
\textit{always} matters. Belief is not easily measured with high accuracy. 
The extent of approximation hidden in the prior densities is seldom considered 
in Bayesian analyses. What is sometimes suggested is to perform a robustness
analysis~{\cite{bayesbook}} (sensitivity analysis of the posterior by
considering, individually, the effects of a small number of potential
alternative choices of a model component, such as parameterization or prior distribution).
However, this concept is not well-defined. 
It lacks a criterion of what is an acceptable change of the posterior and also 
which class of priors should be used.

It is often stated that the data swamp the prior: the prior is washed out as
the number of observation increases. The statement of the ``washing out'' of
the prior lacks a qualitative and also quantitative proof and would need to be
verified case-by-case. Furthermore, the possibility of eventual convergence of belief is 
irrelevant to the day-to-day problem of learning from data in science.
Moreover, it provides only an illusion of the existence of an objective probability~\cite{probaphilo1} 
(the eventual convergence of opinions by remaining coherent (internal consistency with the probability axioms) 
is not enough to guarantee that the Bayesian answer is a good answer to a real-world question). 

We cannot verify if the Bayesian probability $P (\theta | \{\mathbf{x}\})$ is
``correct'' by observing the frequency with which $\theta$ occurs, since this
is not the way Bayesian probability is defined. Hence, it would be odd
trying to justify \textit{post hoc} the priors on frequentist grounds~{\cite{Cousins00}}.

\begin{figure}[tb]
\centering
\includegraphics[width=0.47\textwidth]{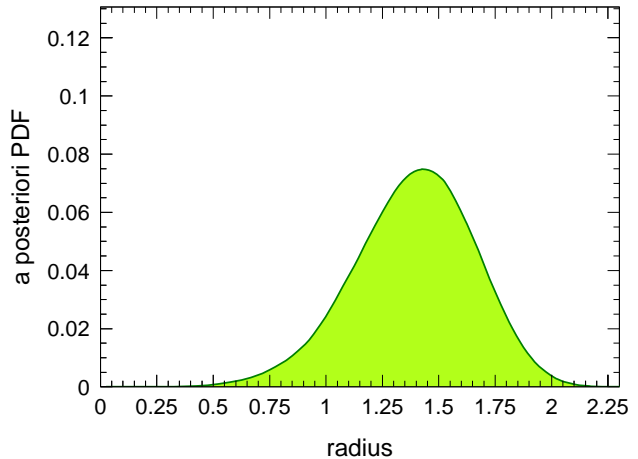}
\vspace{-0.8cm}
\caption{Bayesian posterior \pdf of the radius  $R=\sqrt{\sum_{i=1}^6 x_i^2}$ 
        in a six-dimensional parameter space where no experimental
        input is known for the values of the $x_i$, and where the ``reasonable''
        choice of flat priors for the $x_i$, with $0<x_i<1$, has been used. 
	The Bayesian treatment succeeds to constrain the radius.  
	\label{radius}}
\end{figure}

Let us illustrate with a very simple educated scenario 
how (precise) posterior results can be solely ``determined''
by the multidimensional convolution of prior probability densities.
For instance, we may consider a $N$-dimensional 
parameter space $x_i$, with $i=1,2,... N$, with $0<x_i<1$, in which no
experimental input is known for the values of the $x_i$: one knows
nothing at all. Still, even in such a total absence of knowledge,
a Bayesian treatment can pretend powerful constraints.
If one is interested in the radius $R=\sqrt{\sum_{i=1}^N x_i^2}$ where
lies the true set of model  parameters, the Bayesian answer is
clear. It is shown in Fig.~\ref{radius}, for $N=6$, and for the ``reasonable''
choice of flat priors. One finds $R = 1.39\pm0.27$, where the central value corresponds to the mean radius and 
the errors give the symmetric 68\% posterior probability interval around the mean value.
One has achieved the remarkable feat of learning
something about the radius of the hypersphere, whereas one knew nothing about the Cartesian coordinates and without making any experiment.

%
%

\section{Parameterizations}\label{sec:param}

\begin{table*}[htb]
\caption{
  Ranges taken for the parameters of the various
  parameterizations used in this paper to fit the $B\to \pi \pi$ observables. 
  The ranges for the \MA, PLD and \ES parameterizations are chosen 
  such that they fully contain the posteriors for all parameters (see~Fig.~\ref{fig:posts}). 
  The ranges for the \RI parameterization are chosen similar to the ones in the \MA parameterization.
  The choice of the ranges in the \RI parameterization will be further discussed in Appendix~\ref{appb}. 
  The phases are given in radians.\label{table1}
}
\centering
\vspace{0.2cm}
\setlength{\tabcolsep}{0.6pc}
{\normalsize
\begin{tabular*}{\textwidth}{@{\extracolsep{\fill}}lcc|lcc|lcc|lcc}\hline
\hline
\multicolumn{3}{ c |}{\MA param.} & \multicolumn{3}{c |}{ \RI
param.}& \multicolumn{3}{c |}{ PLD param.} &  \multicolumn{3}{c }{\ES param.} \\
\hline
                   & min.      & max.      &            & min.      &
                   max.    &           & min.      &  max.   &      &
                   min.    &  max. \\
                   & value     & value     &            & value     &
                   value   &           & value     & value   &      &
                   value   &    value \\ 

\hline 
$|T^{+-}|$        &     0     &   10    & 
$|T^{+-}|$        &     0     &   15    & 
$a$               &      0.4  &   1.2   &
$\BR^{+-}(10^{-6})$ &  3 & 7     \\
$|P|$              &      0    &  10    & 
$\mathrm{Re}(P)$   &     $-10$   &   10      & 
$\bar a$           &      0.8  &   1.6     &
$\BR^{+0}(10^{-6})$ &  3 & 8   \\    
$|T^{00}|$          &      0    &   10 & 
$\mathrm{Im}(P)$    &     $-2.5$  &   0     & 
$\mu$               &      1.6  &   2.6     &
$\BR^{00}(10^{-6})$ &  0 & 3    \\    
$\delta_P$ &      0    &$2\pi$ & 
$\mathrm{Re}(T^{00})$ &     $-10$   &   5    & 
$\delta$           &      0    &    $2\pi$ & 
$C^{+-}$ &  $-0.8$ &  0.2     \\           
$\delta_{00}$ &      0    & $2\pi$ & 
$\mathrm{Im}(T^{00})$ &      $-4$   &   4      & 
$\alpha_{\rm eff}$ &      0    &    $2\pi$ &
$S^{+-}$ &  $-1$ & 0  \\     
$\alpha$           &      0    &    $\pi$  & 
$\alpha$   &      0    &  $\pi$  & 
$\alpha$  &       0   &   $\pi$   & 
$C^{00}$ &  $-1$ & 1     \\      
\hline\hline
\end{tabular*}}
\end{table*}

\begin{figure*}[tbhp]
\centering
\includegraphics[scale=0.43]{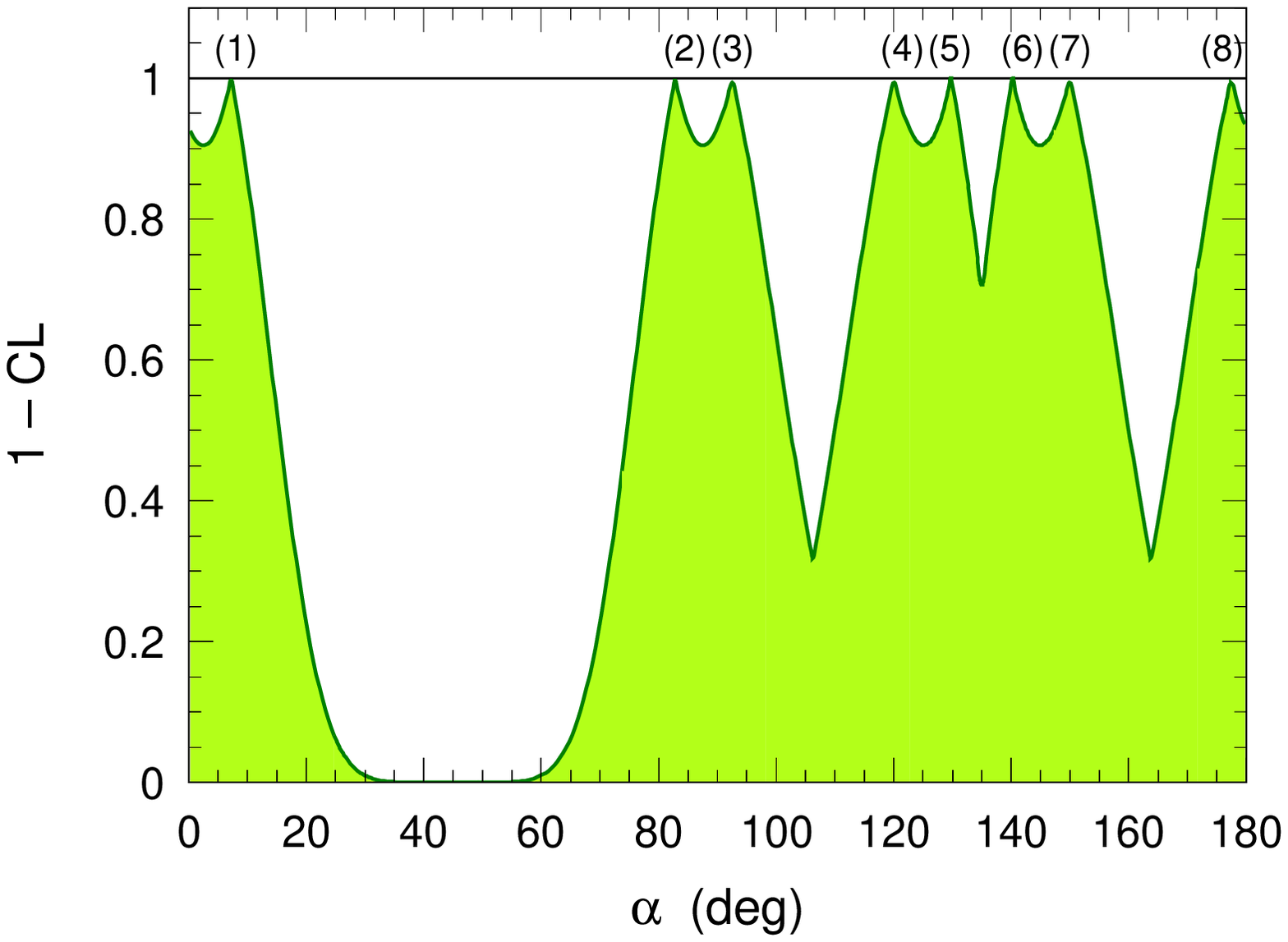} \\
\includegraphics[scale=0.43]{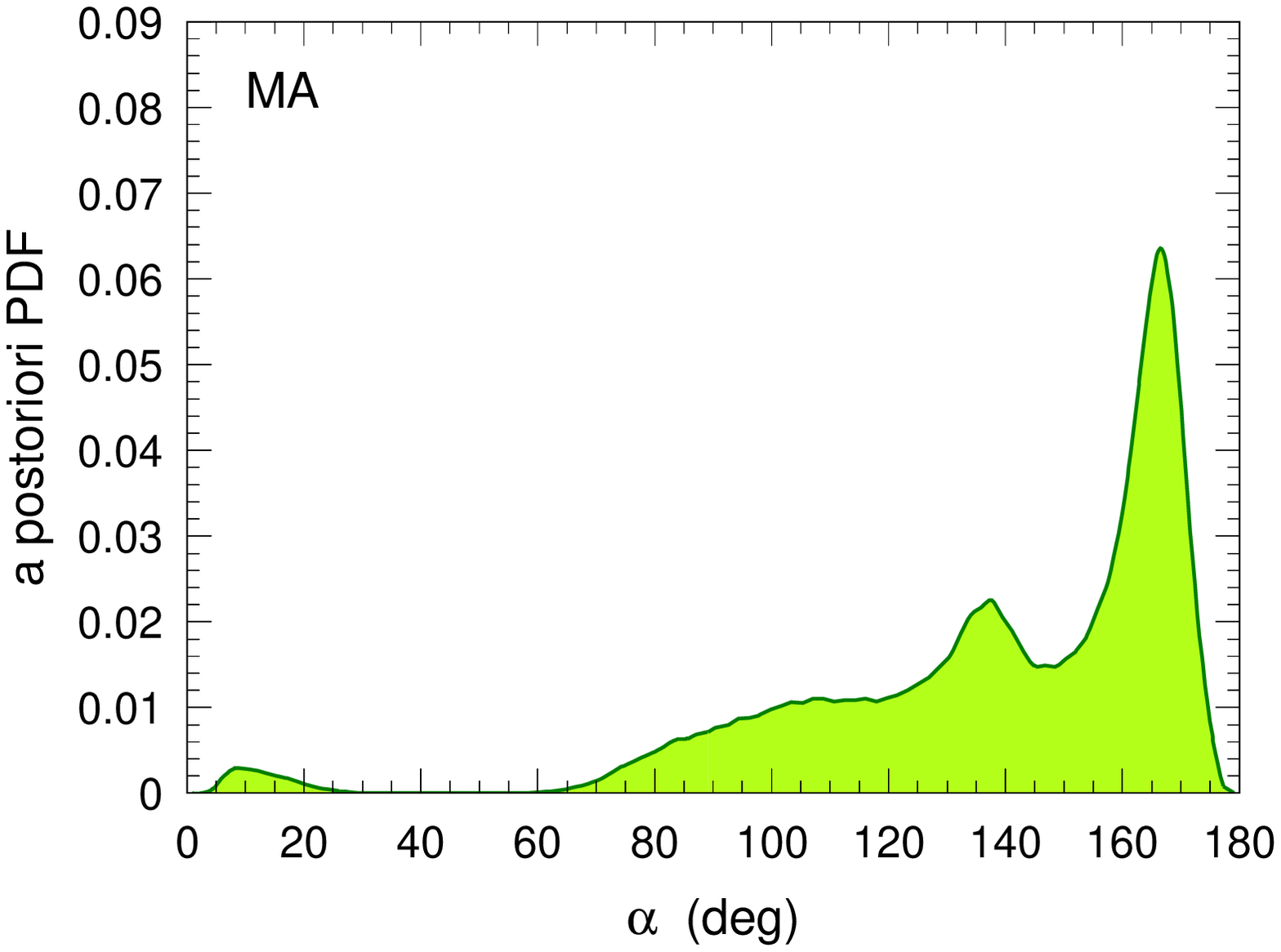}
\includegraphics[scale=0.43]{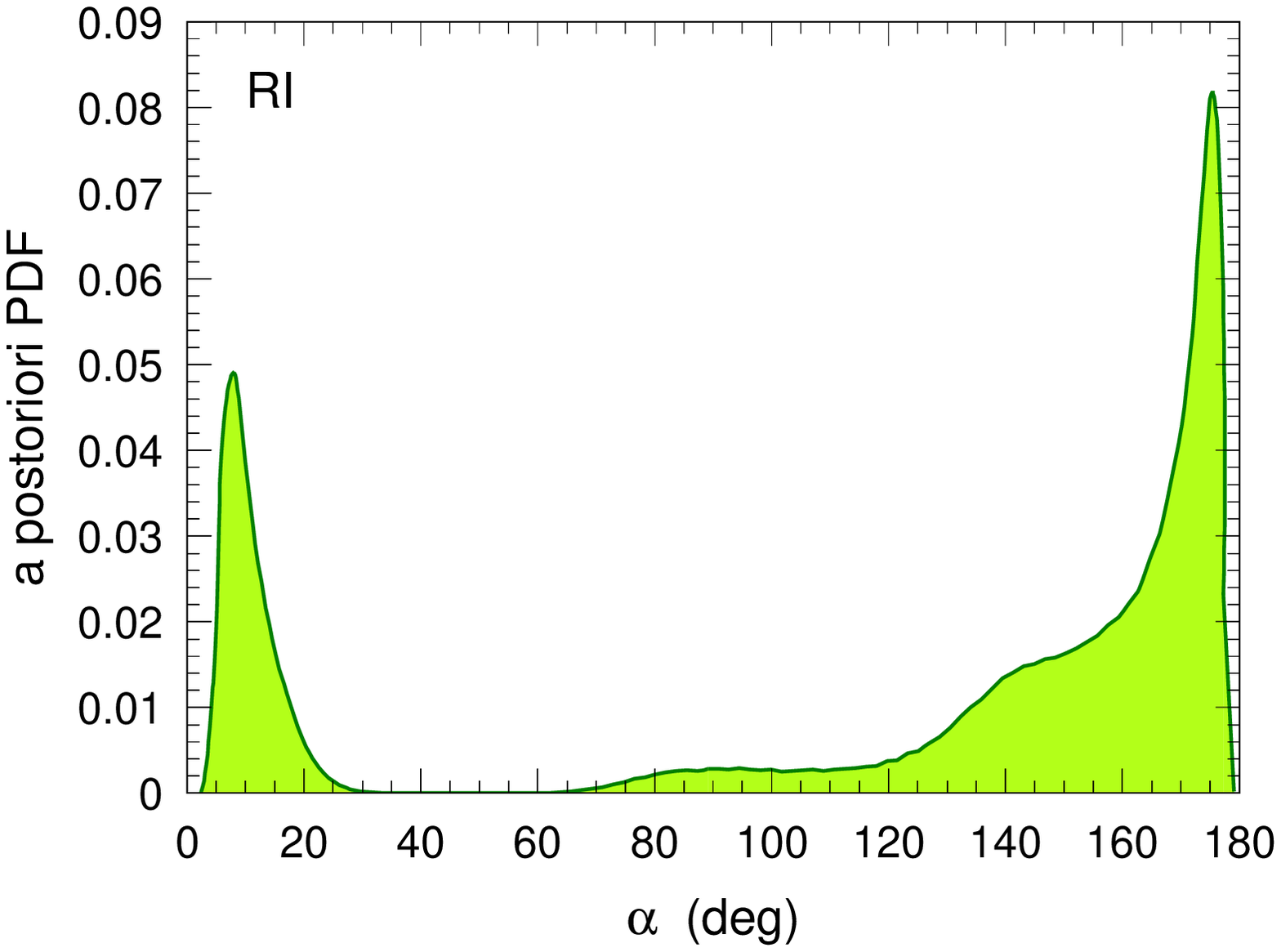}
\includegraphics[scale=0.43]{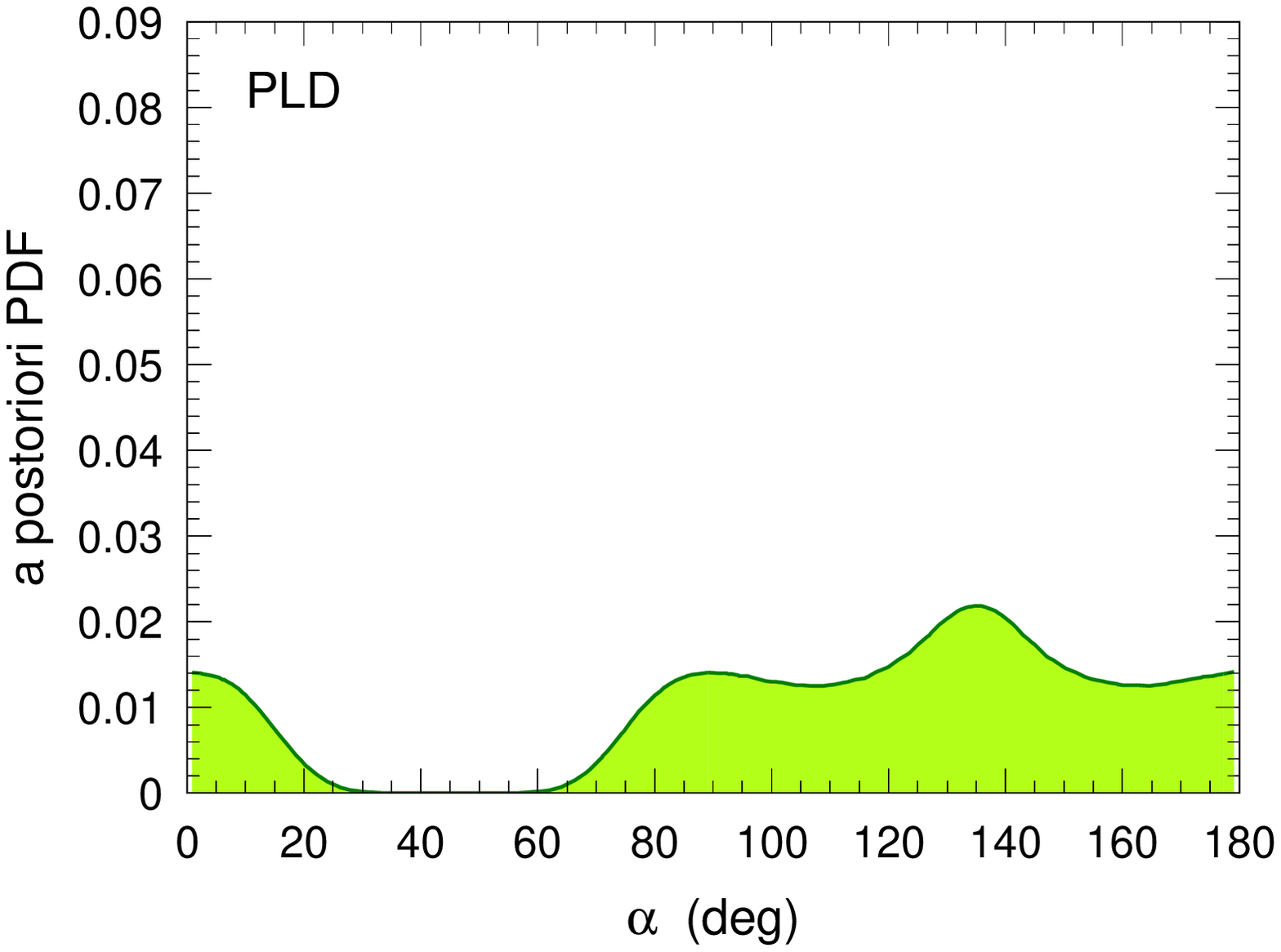}
\includegraphics[scale=0.43]{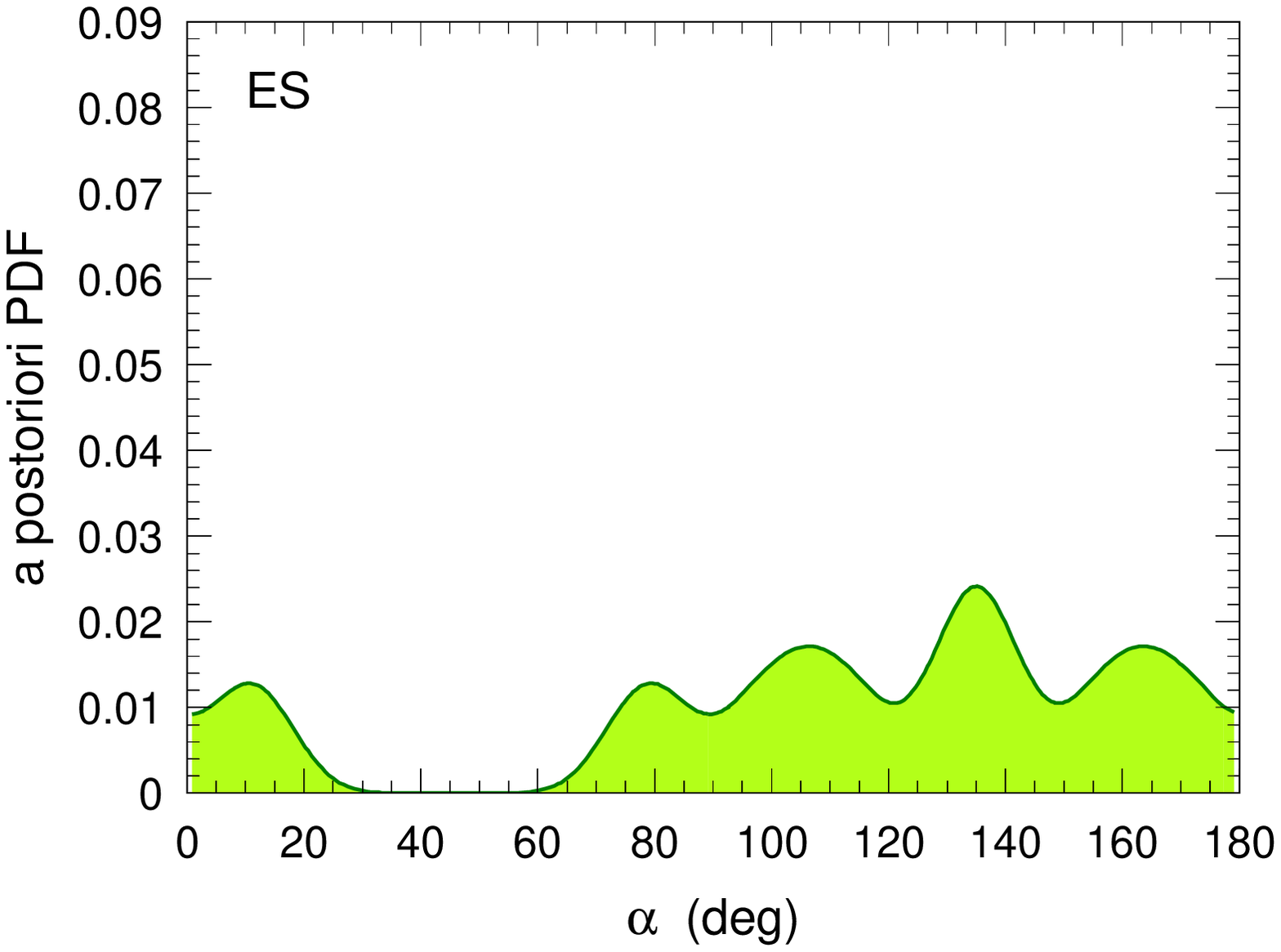}
\vspace{-0.3cm}
\caption{Results for the CKM phase $\alpha$ obtained with the different
         parameterizations from the $\B\to\pi\pi$ isospin analysis. 
         The upper plot shows the frequentist confidence level, which is 
         independent of the parameterization used. The solutions (1) through (8) coincide with 
	 the analytical solutions computed from the expressions given in Footnote~\ref{foot:eight}. The remaining plots
         show the Bayesian a posteriori \pdfs for the parameterizations
         indicated by the labels. The posterior \pdf in the \RI parameterization 
	 depends strongly on the ranges chosen for the priors of $|T^{+-}|$, $\Re(P)$
	 and $\Re(T^{00})$ parameters (see Appendix~\ref{appb}).}
\label{fig1}

\end{figure*}

\begin{figure*}[htbhp]
\centering
\MA parameterization --- a posteriori \pdfs \hspace{2.6cm} \RI parameterization --- a posteriori \pdfs
\includegraphics[width=0.47\textwidth]{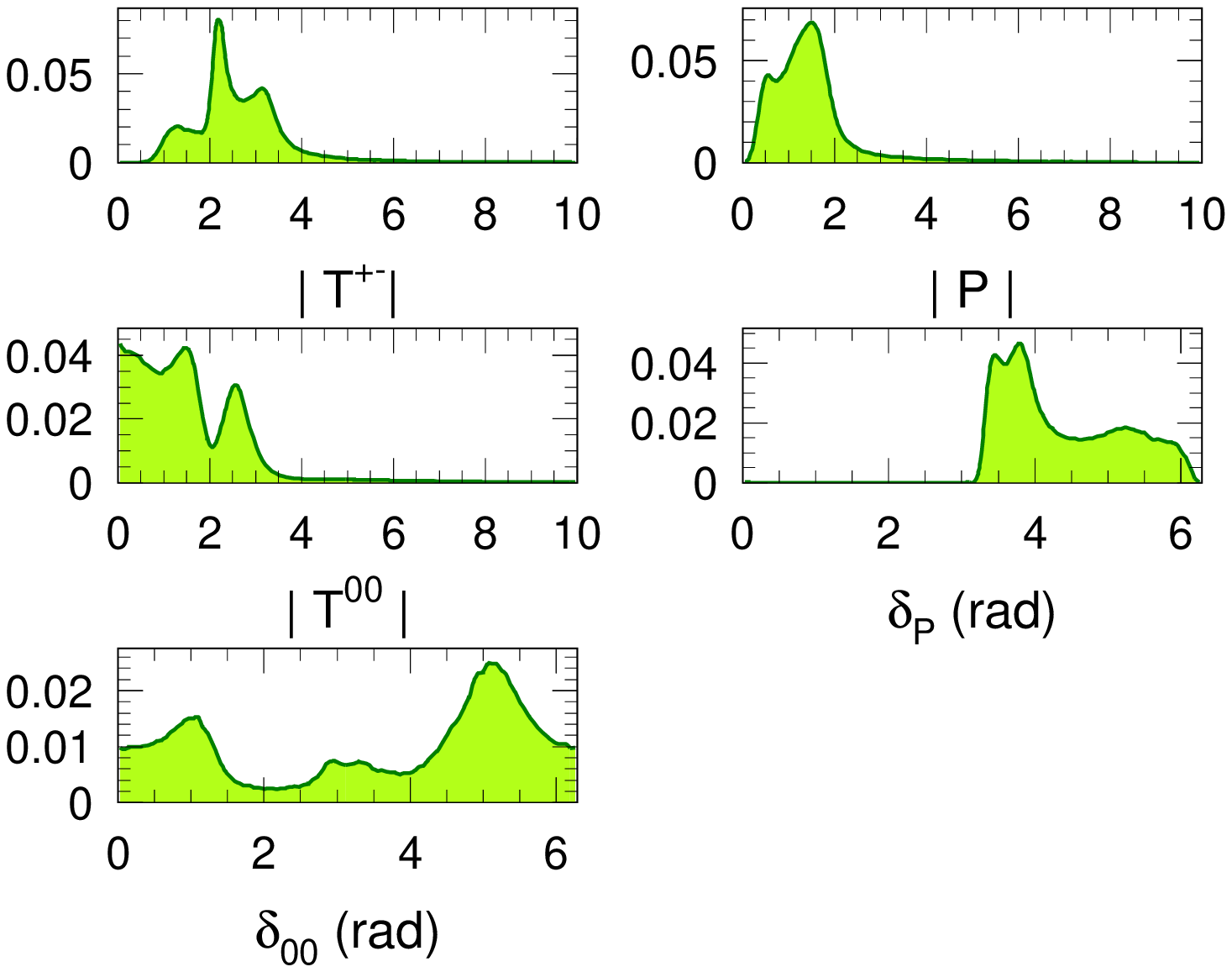}
\includegraphics[width=0.47\textwidth]{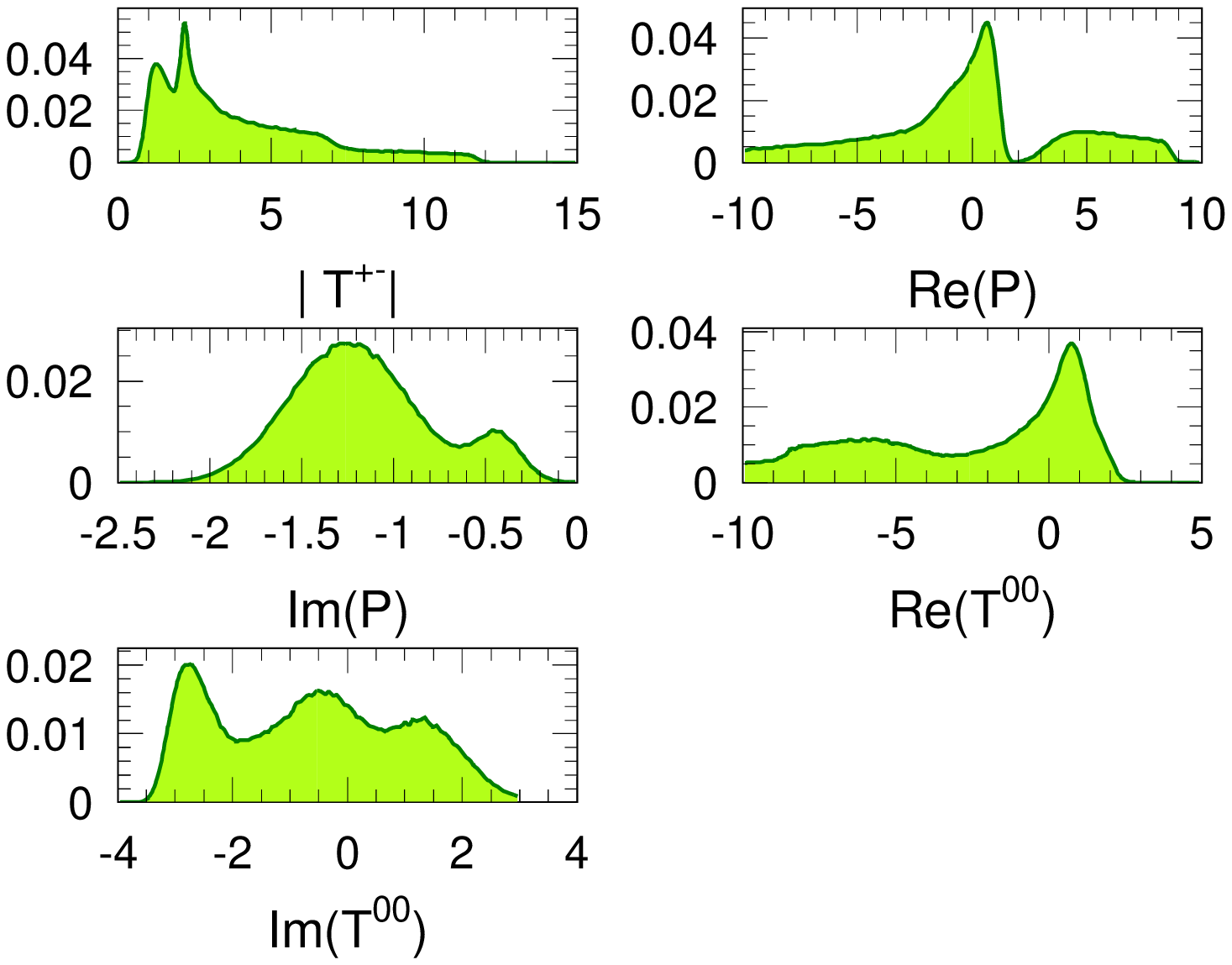}\\
PLD parameterization  --- a posteriori \pdfs \hspace{2.5cm} \ES parameterization  --- a posteriori \pdfs
\includegraphics[width=0.47\textwidth]{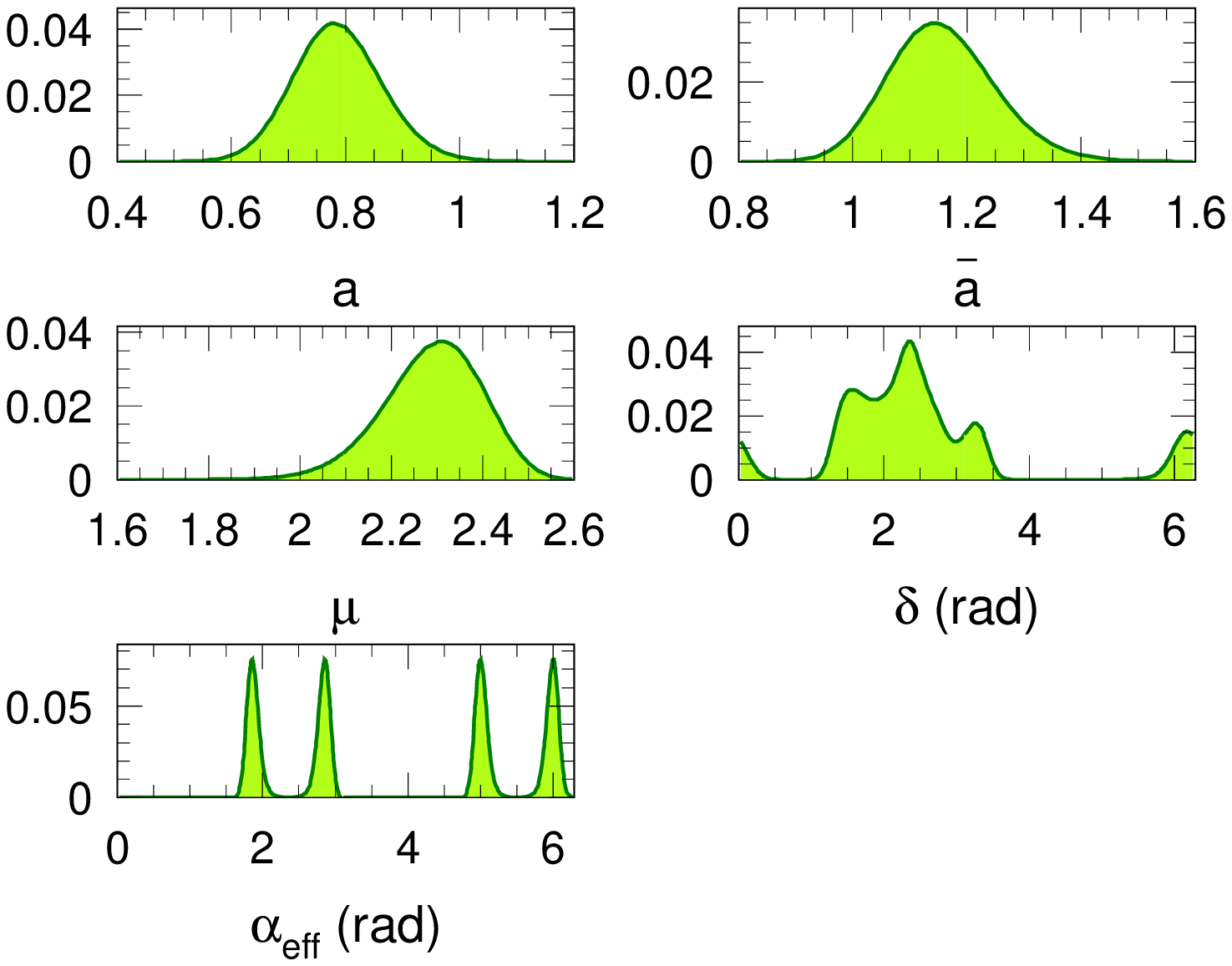}
\includegraphics[width=0.47\textwidth]{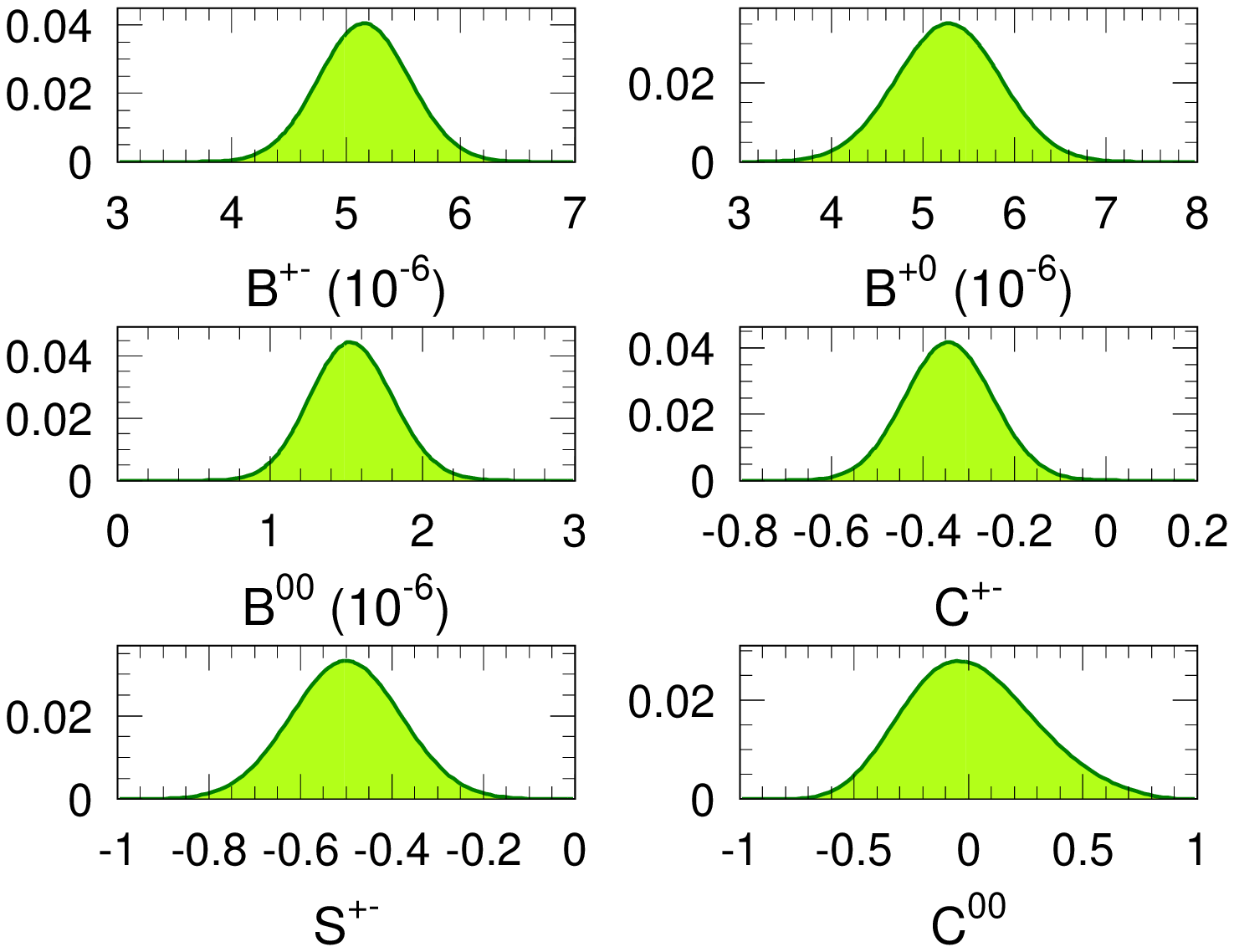}
\vspace{-0.6cm}
\caption{
  Posterior \pdfs for the various model parameters used in this paper except for $\alpha$ which are shown in Fig.~\ref{fig1}.
  One observes that for the \RI parameterization the posterior \pdfs for $\Re(P)$ and $\Re(T^{00})$ 
    do not vanish at the lower edge of their ranges so the ranges should 
    be expanded further. Although the $|T^{+-}|$ posterior appears to be contained in
    its prior range, it turns out that it must also be expanded further. More generally,
    as illustrated by the \RI parameterization, for the sake of consistency, a Bayesian
    treatment must provide the proof that the range chosen for the priors are not introducing
    hidden piece of information. It is shown in Appendix~\ref{appb} that for the \RI parameterization
    the choice of the ranges actually determines the posterior \pdf for $\alpha$.
}\label{fig:posts}
\end{figure*}

In this section we consider the Bayesian treatment applied to 
$B\to\pi\pi$  decays (the Bayesian treatment of $B\to\rho\rho$ decays is 
discussed in Appendix~\ref{apprhorho}). The current world average values for the observables are 
$\BR^{+-} = (5.1\pm0.4)\times 10^{-6}$, 
$\BR^{+0} = (5.5\pm0.6)\times 10^{-6}$, 
$\BR^{00} = (1.45\pm0.29)\times 10^{-6}$, 
$C^{+-} = -0.37\pm0.10$, 
$S^{+-} = -0.50\pm0.12$, 
$C^{00} = -0.28\pm 0.40$ ~\cite{pipi}.
These observables are reproduced by the model with the following eight values 
 for the phase $\alpha$ (in degrees), as computed from the analytical solutions in Footnote~\ref{foot:eight}:
\beq
\{6.3,83.7,91.8,120.7,128.8,141.2,149.3,178.2\}\,. \label{eq:eightsol}
\eeq
We apply the Bayesian treatment  using  the four 
parameterizations in Section~\ref{sec:formalism}:
\begin{itemize}
\item the Standard Model modulus and argument parameterization (\MA);
\item the Standard Model real and imaginary parameterization (\RI);
\item the Pivk-LeDiberder parameterization (PLD);
\item the explicit solution parameterization (\ES).
\end{itemize}
Obviously, one may consider a much larger variety of parameterizations,\footnote
{
  A fifth parameterization is introduced in Appendix~\ref{appb} to discuss further Bayesian peculiarities with the 
   \RI parameterization.
}
but the ones considered here are sufficient to make our point clear.
They are all natural in the sense that they were defined beforehand without
having in mind the present discussion.
For all four parameterizations we use uniform priors for all the 
six model parameters. In particular, the prior \pdf used for $\alpha$ is uniform in the range [$0,\pi$]
for the \MA, \RI and PLD parameterizations (for the \ES parameterization, $\alpha$ is not an input model parameter). 
The choice of uniformity is 
not the result of a strong argument, nor is it particularly natural; rather 
it is taken for the sake of simplicity, and to be conform to the choice made by 
Bayesian analyses already published on the subject~\cite{UTfit1}.
The ranges used for the parameters are given in Table~\ref{table1}. 
The resulting posterior \pdfs for $\alpha$ are shown in Fig.~\ref{fig1}
(the posterior \pdfs for the other model parameters are shown  in Fig.~\ref{fig:posts}).
The top plot gives the $1-{\rm CL}$ result for the 
frequentist treatment. It is independent of the parameterization used.
The eight solutions in Eq.~(\ref{eq:eightsol}) for the phase $\alpha$ are clearly visible,
and correspond exactly to the analytical solutions in Footnote~\ref{foot:eight}.

\subsection{Modulus and Argument Parameterization}

The Bayesian treatment indicates the presence of basically two
mirror solutions; the previous mirror solution at $\alpha\simeq 135^\circ$
and a new solution, which is strongly favored, at $\alpha\simeq 165^\circ$.
One also observes that values nearby $0^\circ$ and $180^\circ$ are 
excluded in this parameterization, in contrast with the following
parameterizations. In the Bayesian approach, all the information 
resides in an individual's posterior probability,
the posterior \pdf for $\alpha$ must be read as the 
individual's updated belief in the plausible values of the parameter $\alpha$.
The individual's degree of belief is thus higher for a value 
of $\alpha\simeq 165^\circ$ than for $\alpha\simeq 135^\circ$.

\subsection{Real and Imaginary Parameterization}

The Bayesian treatment seems to detect the presence of two mirror 
solutions, a mirror solution at $\alpha\simeq 10^\circ$ and another solution, 
which is favored, at $\alpha\simeq 175^\circ$. The posterior \pdf appears to vanish at the 
origin (and at $180^\circ$). However, this 
is only an artifact resulting from the truncation of the prior ranges used for $|T^{+-}|$, 
$\Re(P)$ and $\Re(T^{00})$ (see Footnote~\ref{truncate}). As demonstrated in Appendix~\ref{appb}
the posterior diverges for $\alpha=0^\circ \, (180^\circ$) and is not normalizable.
Expanding the ranges leads to an improper posterior for $\alpha$.

\subsection{Pivk-LeDiberder Parameterization}

The Bayesian treatment only vaguely detects the
presence of mirror solutions. The posterior \pdf tends to favor a value for $\alpha$
($135^\circ$) which corresponds to a dip in the frequentist $1-{\rm CL}$. 
This is a hint that the Bayesian 
treatment introduces a piece of information which is ``missed'' by the
frequentist analysis. Since the latter uses all the available
experimental data, this additional piece of information must be
embedded in the priors. One also observes that values nearby $0^\circ$
and $180^\circ$ are not disfavored.

\subsection{Explicit Solution Parameterization}

The Bayesian treatment detects the presence of mirror solutions, but 
like in the PLD parameterization, one solution is favored for $\alpha$ ($135^\circ$).
The fact that one solution is favored over the others is because two nearby mirror solutions are 
superimposed in the Bayesian treatment (see Section~\ref{sec:mirror2D}).
This feature is present in all parameterizations. In the \RI parametrization, it is reflected 
only as a shoulder in the posterior \pdf.
One also observes that values nearby $0^\circ$ and $180^\circ$ are not disfavored. 
The posterior \pdf for $\alpha$ is akin to the one obtained with the PLD parameterization. 
This is because the latter parameterization is chosen close to the 
measured quantities.

\subsection{Conclusion}

The Bayesian \pdfs result from a prior- and parameterization-dependent
weighted average of data with mirror solutions. The fact that the posterior \pdfs
in the \MA and \RI parameterizations are so different compared to
the PLD and \ES parameterizations is due to a strong prior dependence.
The behavior of the posterior \pdf at the origin (and at $180^\circ$) also strongly depends 
on the parameterization (see Appendix~\ref{app:alimit}).

From all these results, what is a scientist supposed to communicate as 
a value for the parameter $\alpha$, without forgetting that they are all based on the same data?
Bayesianism is simply a system for keeping one's internal beliefs self-consistent. 
It is not concerned with whether or not these beliefs represent the information content of the data.

%
%

\section{Removing Essential Information}

\begin{figure*}[htb]
\centering
\includegraphics[scale=0.43]{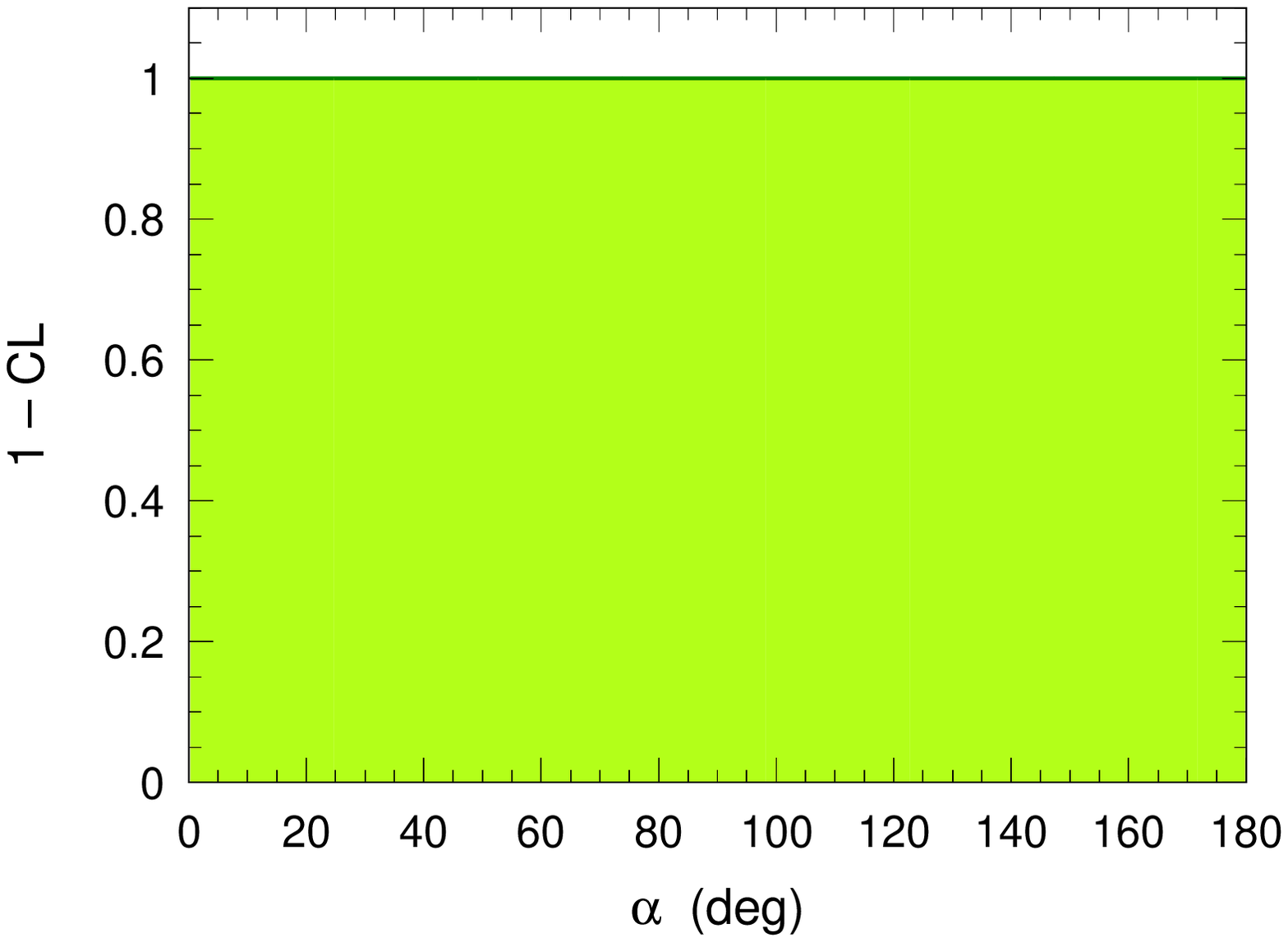} \\
\includegraphics[scale=0.43]{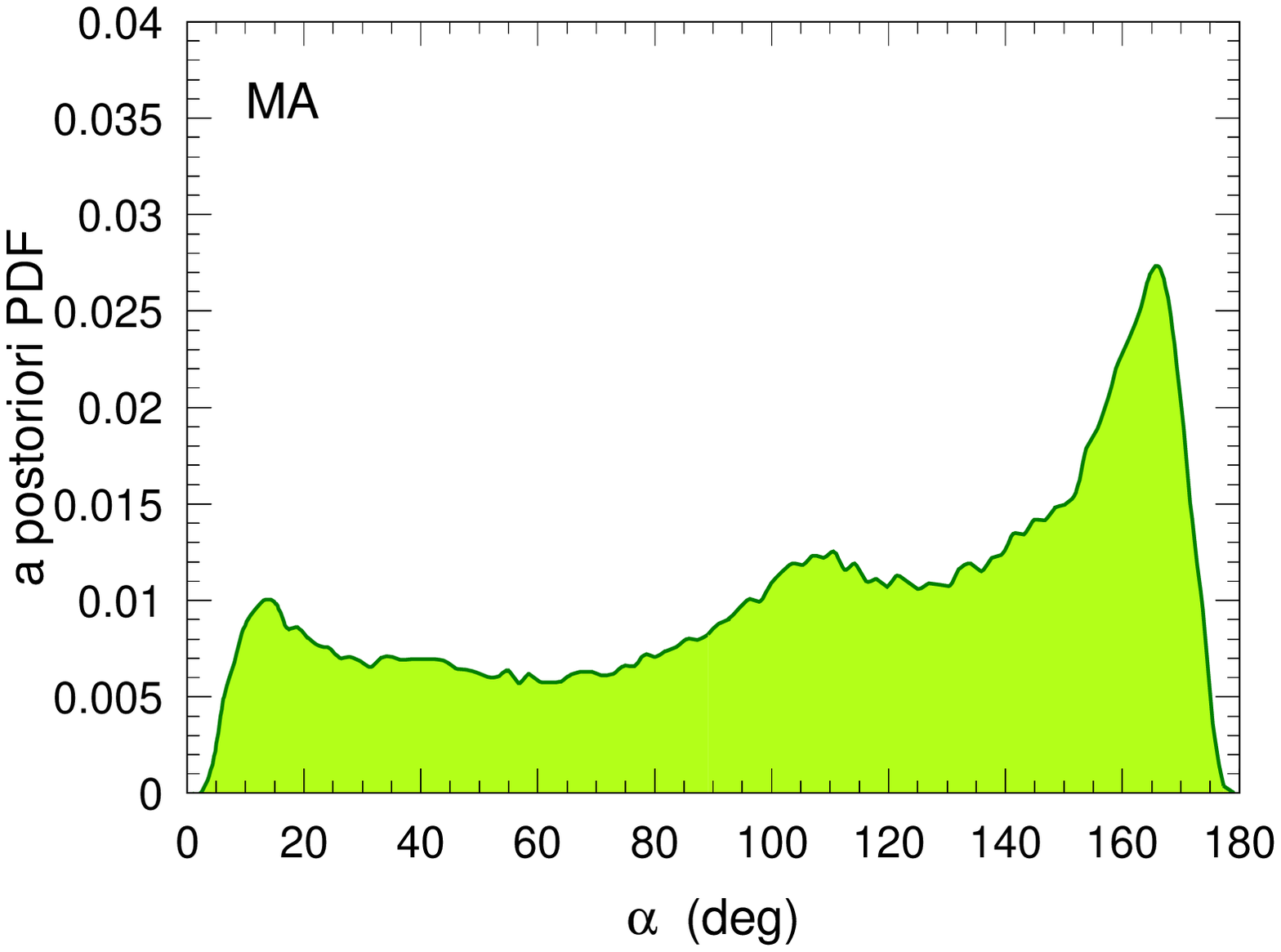}
\includegraphics[scale=0.43]{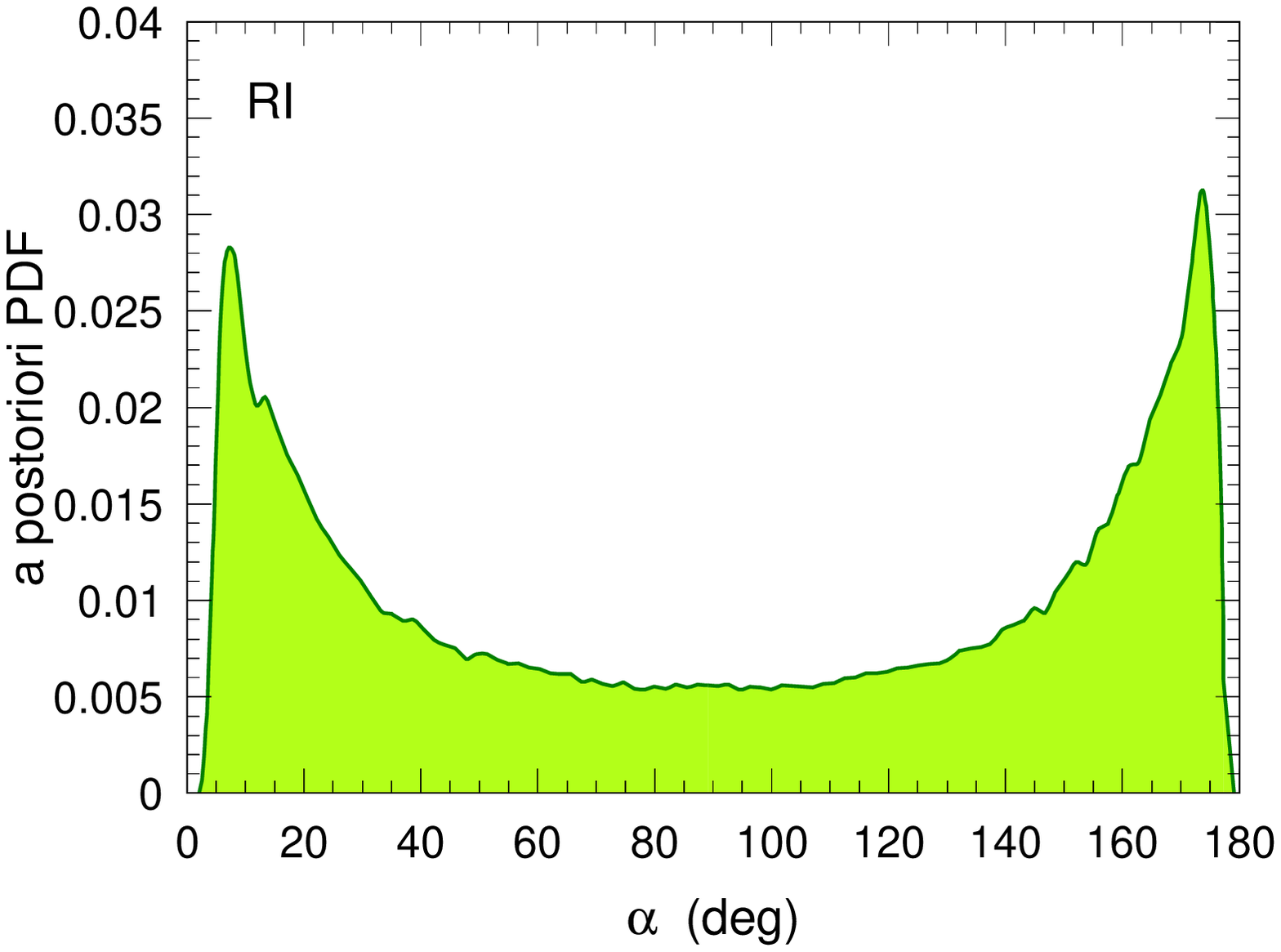}
\includegraphics[scale=0.43]{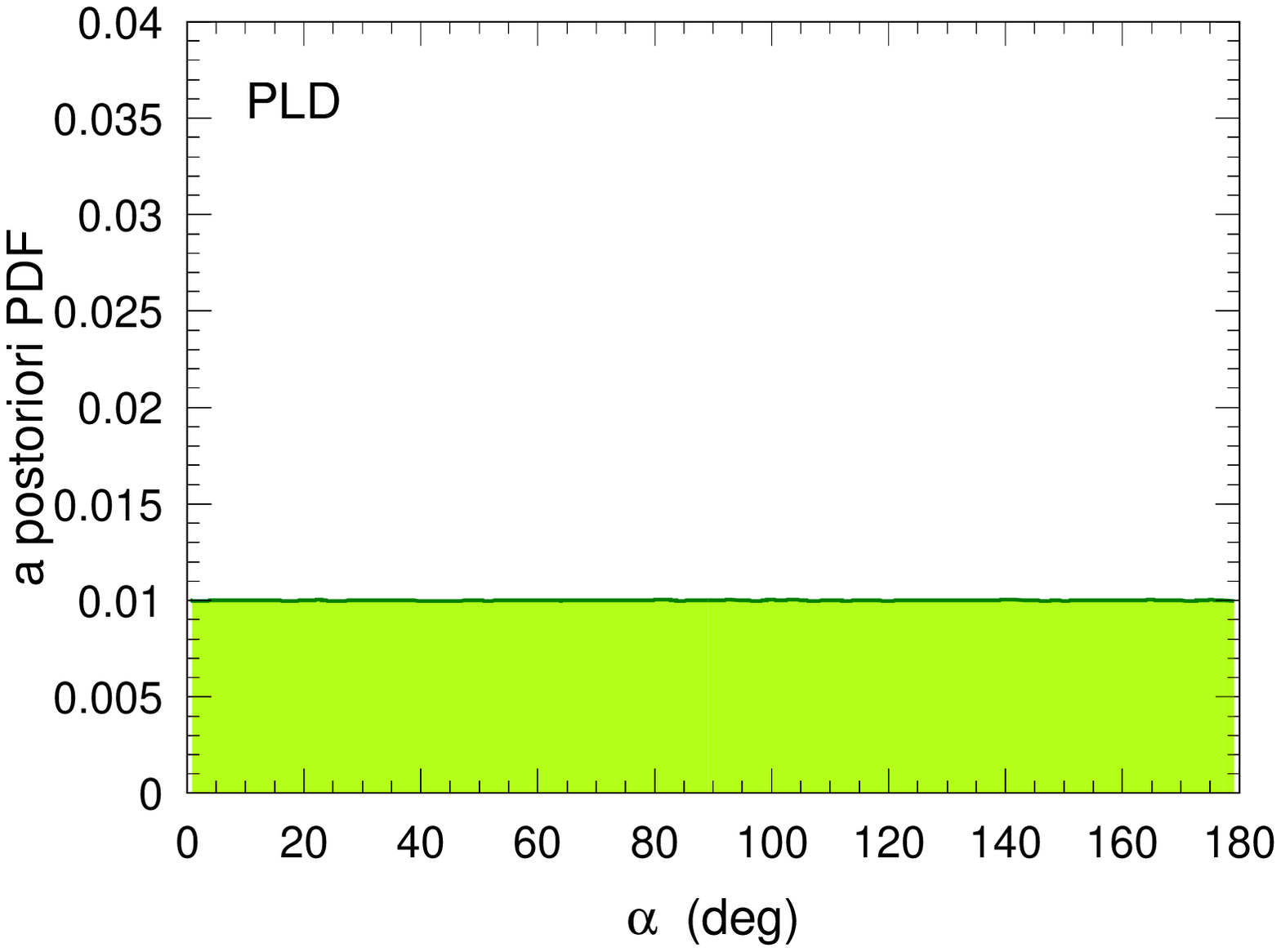}
\includegraphics[scale=0.43]{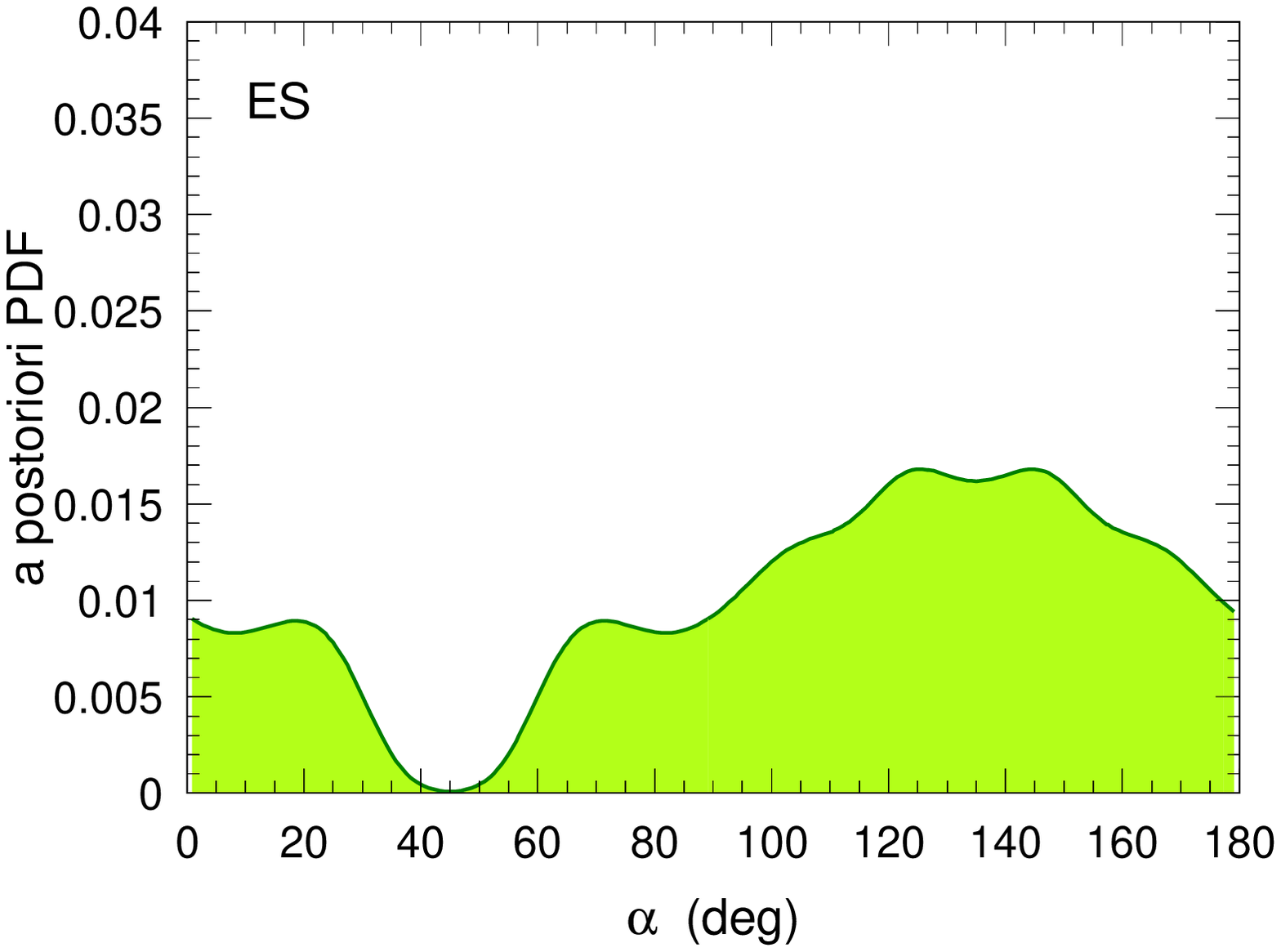}
\vspace{-0.3cm}
\caption{Results for the CKM phase $\alpha$ obtained with the different
         parameterizations from the $\B\to\pi\pi$ isospin analysis,
	 without using the input from the $\Bz\to\piz\piz$ branching fraction and 
         direct \CP-asymmetry measurements. No model-independent
	 constraint on $\alpha$ can be inferred in this case (with the exception of
	 the exclusion of the singular points $\alpha=0,\pi$). 
         The upper plot shows the frequentist confidence level, which is 
         independent of the parameterization used. The remaining plots
         show the Bayesian a posteriori \pdfs for the parameterizations
         indicated by the labels (the unsmooth aspect of some \pdfs is 
         due to the fact that we used ``only'' $10^{10}$ (sic!) Monte Carlo
         events to carry out the numerical integration). }
\label{fignoinfo}
\end{figure*}
It is instructive to study the behavior of the Bayesian posteriors in
a situation where one deliberately removes crucial data from the analysis, \eg, 
when performing the fit without using the $\Bz \to \pi^0 \pi^0$ branching fraction and 
direct \CP-asymmetry measurements. In this case it is well known~\cite{babarphysbook} 
that no information on $\alpha$ can be derived from the data (with the exception 
of the exclusion of $\alpha=0$ in case of non-zero \CP violation, see Appendix~\ref{appalpha}). In effect, having 
carried out the fit for a given $\alpha$ value, the values of the model
parameters that correspond to this fit can be used to compute explicitly the 
values of the model parameters corresponding to any other value of $\alpha$, 
if non-zero (see Appendix~\ref{appalpha}): this second set of values yields 
a fit of exactly the same quality ($\chi^2$). Stated differently, the posterior 
for $\alpha$ provided by the Bayesian treatment, if unbiased, must
be uniform since data do not favor any value of $\alpha$. The posterior \pdf obtained 
for the four parameterizations are shown in  Fig.~\ref{fignoinfo}. While the PLD 
parameterization yields the expected uniform \pdf, the three others do not: they 
are able to extract information on $\alpha$, which is introduced by the priors.

%
%

\section{Mirror solutions in a simple 2D problem}
\label{sec:mirror2D}

In this section we present a simple and solvable two-dimensional example to illustrate
how mirror solutions make the Bayesian
approach fail. We work within a theory that predicts the expressions of
two observables $X$ and $Y$ as a function of the two parameters $\alpha$ and
$\mu$
\begin{eqnarray}
X &=& (\alpha+\mu)^2 \,,\nonumber\\
Y &=& \mu^2\,.\label{2Dtheory}
\end{eqnarray}
Assuming $(X_m,Y_m)$ are measured values for the observables, the central values
for the parameters can be found by inverting the above system
\begin{eqnarray}
\alpha_0 &=& \epsilon_X\, \sqrt{X_m} -\epsilon_Y\, \sqrt{Y_m}\,,\label{exact2D}\\
\mu_0 &=& \epsilon_Y\, \sqrt{Y_m} \,, \nonumber
\end{eqnarray}
where $\epsilon_X,\,\epsilon_Y=\pm 1$.
Hence there are in general four solutions for $(\alpha_0,\mu_0)$ for a
given set of measurements. Note that this example is far from being
academic, since the theoretical expressions above are very similar to the
usual amplitudes $\leftrightarrow$ branching fraction relations in particle
physics. The pattern of discrete ambiguities is also very similar to the
one encountered in the isospin analysis for the CKM phase $\alpha$.

\begin{figure}[tb]
\includegraphics[width=0.47\textwidth]{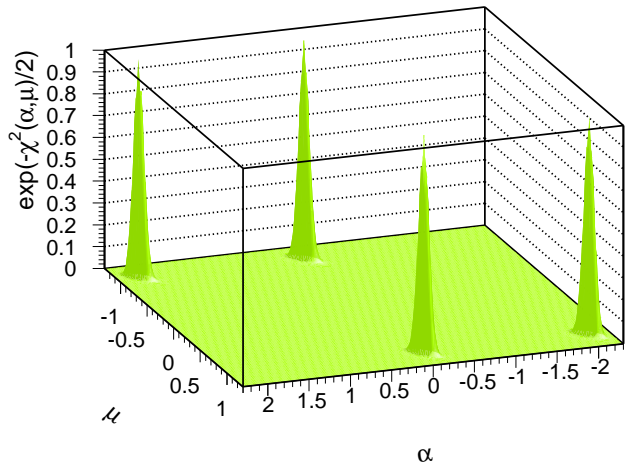}
\vspace{-0.6cm}
\caption{2D example (Eqs.~(\ref{2Dtheory}) and (\ref{2Ddata})): the
values of $\exp(-\chi^2/2)$ as a function of the parameters 
$(\alpha,\mu)$.}\label{2DPDF}
\end{figure}

If $\alpha$ and $\mu$ are fundamental physics parameters, Nature can only
accommodate a single pair of values. This means that the
representation of the four-valued discrete ambiguity
$(\epsilon_X,\,\epsilon_Y)=(+1,+1)\,,(+1,-1)\,,(-1,+1)\,,(-1,-1)$ must be
interpreted as a logical exclusive OR operator.

\begin{figure}[tb]
\includegraphics[width=0.47\textwidth]{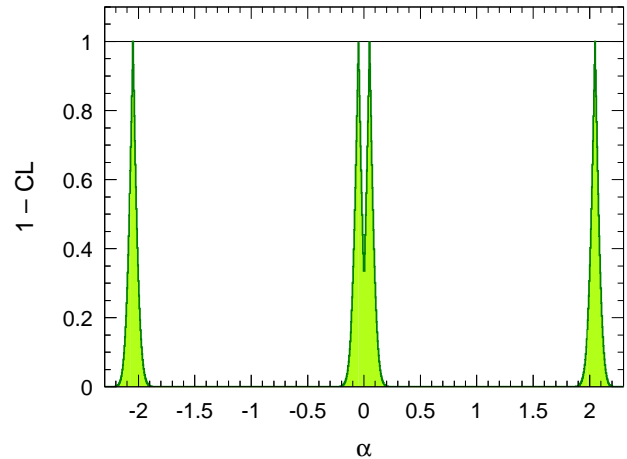}
\includegraphics[width=0.47\textwidth]{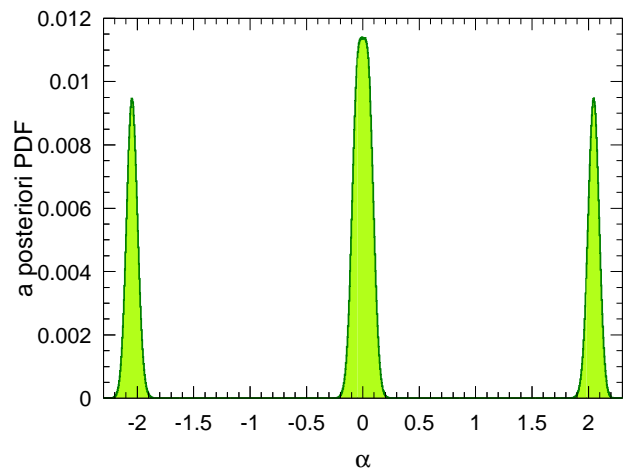}
\vspace{-0.4cm}
\caption
{
  Top: result of the frequentist fit to the data. The peaks are exactly located at the 
  analytical solutions $\alpha = -2.05$, $-0.05$, $0.05$, $2.05$ computed from 
  Eq.~(\ref{exact2D}).  Bottom: a posteriori \pdf according to the Bayesian treatment. The most probable value is biased.
}
\label{2Dfit}.
\end{figure}

We assume that an experiment has measured the observables from a Gaussian
sample of events, with the results
\begin{eqnarray}
X &=& 1.00 \pm 0.07\,,\nonumber\\
Y &=& 1.10 \pm 0.07\,.\label{2Ddata}
\end{eqnarray}
While the measurement is reasonably precise (below the 10\% level), and the mirror
solutions are well separated in the $(\alpha,\mu)$ space (see
Fig.~\ref{2DPDF}),
the central values correspond to a somewhat ``unlucky'' situation since,
as shown in Fig.~\ref{2Dfit},
the two solutions for small $\alpha$ overlap somewhat. This kind of overlapping
precisely corresponds to what may occur in the isospin analysis of the
$B\to\pi\pi$ data.

The result of the Bayesian procedure applied to this 2D example is shown
in Fig.~\ref{2Dfit}. One immediately sees the striking difference with
respect to the frequentist fit: after the marginalization with respect to
the nuisance parameter $\mu$, only one peak instead of two is left close to the origin in the
$\alpha$ constraint, and its best value is biased with respect to
the minimum $\chi^2$ ones (which coincide with the explicit
solution of Eq.~(\ref{exact2D})).

This unexpected behavior is best understood by looking at
Fig.~\ref{2DPDF}. In this space the likelihood is Gaussian and the
four solutions do not overlap; while the minimum-$\chi^2$ fit
selects, for each value of $\alpha$, the best value of $\mu$ with respect
to the data, the Bayesian procedure \textit{integrates}~\cite{BonaFLHC}
all events in the $\mu$ direction that correspond to the same value of
$\alpha$. In other words, the frequentist approach naturally implements
the logical exclusive OR operating on the solutions, in contrast to the
Bayesian approach that effectively replaces OR by AND. The latter is
clearly unacceptable if one is used to the common wisdom that fundamental
parameters have a definite single value realized in Nature. In the present
case, the counter-intuitive result of the Bayesian procedure is that the
most probable  value of $\alpha$ is $\alpha=0$, which in turn implies
$X=Y$ that is not the situation preferred by the data (see Eq.~(\ref{2Ddata})).

%
%

\section{Origin of the problem}

The origin of the problem lies in the very first Bayesian assumption,
namely that unknown model  parameters are to be understood as
mathematical objects distributed according to \pdfs, which
are assumed to be known: the priors. Obviously, the choice
of the priors cannot be irrelevant; hence, the Bayesian treatment is doomed
to lead to results which depend on the decisions made, necessarily on
unscientific basis, by the authors of a given analysis, for the choice
of these extraordinary \pdfs. This is not a new situation. 
This limitation --- deliberately mixing information coming from
scientific data together with human inputs --- is frequently perceived as mild, 
because ``reasonable variations'' of the human inputs lead to ``reasonable variations''
of the output of the Bayesian treatment. Moreover, a particular shape
of \pdf is declared to be ``reasonable'': the uniform \pdf. This is
because, implicitly, one feels that using a flat distribution for
priors is akin to using no information at all, hence injecting only a
weak a priori assumption through the priors.

What ``reasonable variations'' means is no more no less clear as
what ``systematics uncertainties''~\cite{systematics} mean in general: 
the dependence on the priors can be viewed as a systematic uncertainty of human origin. 
What is particular in the Bayesian treatment discussed in this article is that 
``reasonable variations'' of the human inputs lead to clearly
``unreasonable variations'' of the output of the Bayesian treatments.
For example, the posterior \pdf for $\alpha$ (see Fig.~\ref{fig1}) using 
the \RI parameterization (with the ranges of Table~\ref{table1}) 
leads one to conclude that the Standard Model is close to being
defeated; knowing that the bulk of data (from other \B decays) 
indicates a value of the phase $\alpha = (100^{+15}_{-9})^\circ$~\cite{jerome:fpcp06}.
Worse, if one widens to infinity the ranges used for the \RI parameterization,
the Bayesian treatment fails to provide a proper \pdf for $\alpha$ (see appendix~\ref{appb}).
The key here is that the statistical analysis dealt with is performed
in a multivariate parameter space. It is well known that the naive use
of ``uninformative'' priors, especially in multivariate spaces, suffer
from various paradoxes and is thus widely criticized in the
literature~\cite{Kass, Seidenfeld}. Our examples are even more striking
because of strong non-linearities and exact degeneracy among several
mirror solutions, and show that the analysis presented in Ref.~\cite{UTfit1} bears no physical meaning.

%
%

\section{Conclusion}

We have demonstrated in this paper that in the isospin analysis of $\B\to\pi\pi$ decays used to
constrain the CKM phase $\alpha$, despite the rather precise experimental inputs,
the posterior \pdf for $\alpha$ shows a striking
dependence on shape and ranges used for the prior \pdfs, {\em and} on the 
choice of the amplitude parameterization. 
The very frequent naive use of flat priors  as presumed
``non-informative'' priors  hides important  unwarranted
assumptions, which may easily invalidate the analysis. 
As stressed by David R. Cox at the Phystat05 conference~{\cite{cox}}: 
``It seems to be agreed from all theoretical standpoints that flat or 
ignorance priors are dangerous, although they are widely used [in Bayesian statistics]. 
Flat priors in several dimensions may produce clearly unacceptable answers.''
Flat priors {\em are} informative. There is a big difference between no
knowledge, and a distribution that is uniform in some (arbitrary)
parameterization. Contrary to what is often stated, Bayesian
and frequentist statistics can lead to very different conclusions
about the same data, even in the limit of a large data sample. The
choice of priors \textit{always} matters in the Bayesian treatment.

The proponents for the use of Bayesian statistics in High Energy
Physics often argue that "everybody is Bayesian in everyday life".
It is certain that Bayesian reasoning dominates the decision making
mechanism of our everyday life. Scientific results however differ 
in many aspects from everyday-life reasoning. In particular, the 
translation of prior beliefs into a robust posterior number, 
representing universal knowledge is impossible. Only the exact 
knowledge of all prior assumptions (including the parameterization, 
\pdf shapes and parameter ranges) used in a certain Bayesian analysis 
allows another one to reproduce its posterior result. Bayesian 
statistics has been shown to be useful for decision making, such as 
buying and selling stock options, or filtering out unwanted electronic 
mails. The decision-making problem is not scientific in nature.

\begin{table*}[tbh]
\caption{Ranges taken for the parameters of the various
   parameterizations used to fit the $B \to \rho \rho$ observables. The phases are given in radians.
   \label{table2}}
\centering
\vspace{0.2cm}
\setlength{\tabcolsep}{0.6pc}
{\normalsize
\begin{tabular*}{\textwidth}{@{\extracolsep{\fill}}lcc|lcc|lcc|lcc}\hline
\hline
\multicolumn{3}{ c |}{\MA param.} & \multicolumn{3}{c |}{ \RI
param.}& \multicolumn{3}{c |}{ PLD param.} & \multicolumn{3}{c}{\ES  param.} \\
\hline
                   & min.      & max.      &            & min.      &
                   max.    &           & min.      &  max.  & & 
                   min. & max. \\
                   & value     & value     &            & value     &
                   value   &           & value     & value   &  &
		   value & value \\

\hline 
$|T^{+-}|$        &      0    &  10    & 
$|T^{+-}|$       &     0     &   40   & 
$a$                &      0.8  &   2.4     & 
$\BR^{+-}(10^{-6})$ &  10 & 45     \\
$|P|$        &      0    &  10    & 
$\mathrm{Re}(P)$   &     $-35$   &   35      & 
$\bar a$           &      0.8  &   2.4     & 
$\BR^{+0}(10^{-6})$ &  0 & 30 \\
$|T^{00}|$      &      0  &   10    & 
$\mathrm{Im}(P)$   &     $-8$  &   8     & 
$\mu$              &      2.2  &   4.5     & 
$\BR^{00}(10^{-6})$ &  0 & 2.5  \\ 
$\delta_P$ &      0    &  $2\pi$ & 
$\mathrm{Re}(T^{00})$ &     $-40$   &   10    & 
$\delta$           &      0    &    $2\pi$ & 
$C^{+-}$ &  $-0.6$ & 0.6     \\           
$\delta_{00}$ &      0    & $2\pi$ & 
$\mathrm{Im}(T^{00})$ &      $-8$   &   8    & 
$\alpha_{\rm eff}$ &      0    &    $2\pi$ & 
$S^{+-}$ &  $-1$ & 1 \\     
$\alpha$           &      0    &    $\pi$  & 
$\alpha$   &      0    &  $\pi$  & 
$\alpha$  &       0   &   $\pi$   & 
$C^{00}$ &  $-1$ & 1     \\      
\hline\hline
\end{tabular*}}
\end{table*}

Leonard J. Savage, a founder of modern subjective Bayesianism makes
 very clear throughout his work that the theory of personal
 probability ``is a code of consistency for the person applying it, not
 a system of predictions about the world around him''. 
``But is a personal code of consistency, requiring the quantification
of personal opinions, however vague or ill formed, an appropriate
basis for scientific inference?''~\cite{Mayo}.
Isn't a hypothesis in physical science either true or false? What is hence
the meaning of the prior in this case? 
The Bayesian personal probability theory misses the main point: ``it confuses feeling with fact''~\cite{probaphilo2}.
In physical sciences, the theory is intended to describe the external physical universe, 
rather than one's internal psychological state. 

Science has to summarize the available information the best it can.
The Bayesian tools do not tell us what we want to know in science. 
What we seek are methods for generating and analyzing data and 
for using data to learn about experimental processes in a reliable manner. 
The kinds of tools needed to do this are crucially different from those the Bayesian statistics supplies~\cite{Mayo}.
These difficulties with the personalistic Bayesian approach indicate that there is a need for 
objective probabilities and a methodology for statistics based on testing.
However the test statistics is chosen 
allows for an objective interpretation of the scientific results~\cite{CoxMayo}. 

%
%

\begin{acknowledgments}
Centre de Physique Th\'eorique is UMR 6207 du CNRS associ\'ee aux
Universit\'es d'Aix-Marseille I et
II et Universit\'e du Sud Toulon-Var; laboratoire affili\'e \`a
la FRUMAM-FR2291. Work partially supported by EC-Contract HPRN-CT-2002-00311 (EURIDICE).
A preliminary version of this work has been presented at the PHYSTAT05 conference. 
Special thanks to the organizers of this stimulating conference.
\end{acknowledgments}
%
%
%

\appendix

\section{Bayesian treatment applied to \boldmath $B\to \rho \rho$ decays}
\label{apprhorho}

\begin{figure*}[tbhp]
\centering
\includegraphics[scale=0.43]{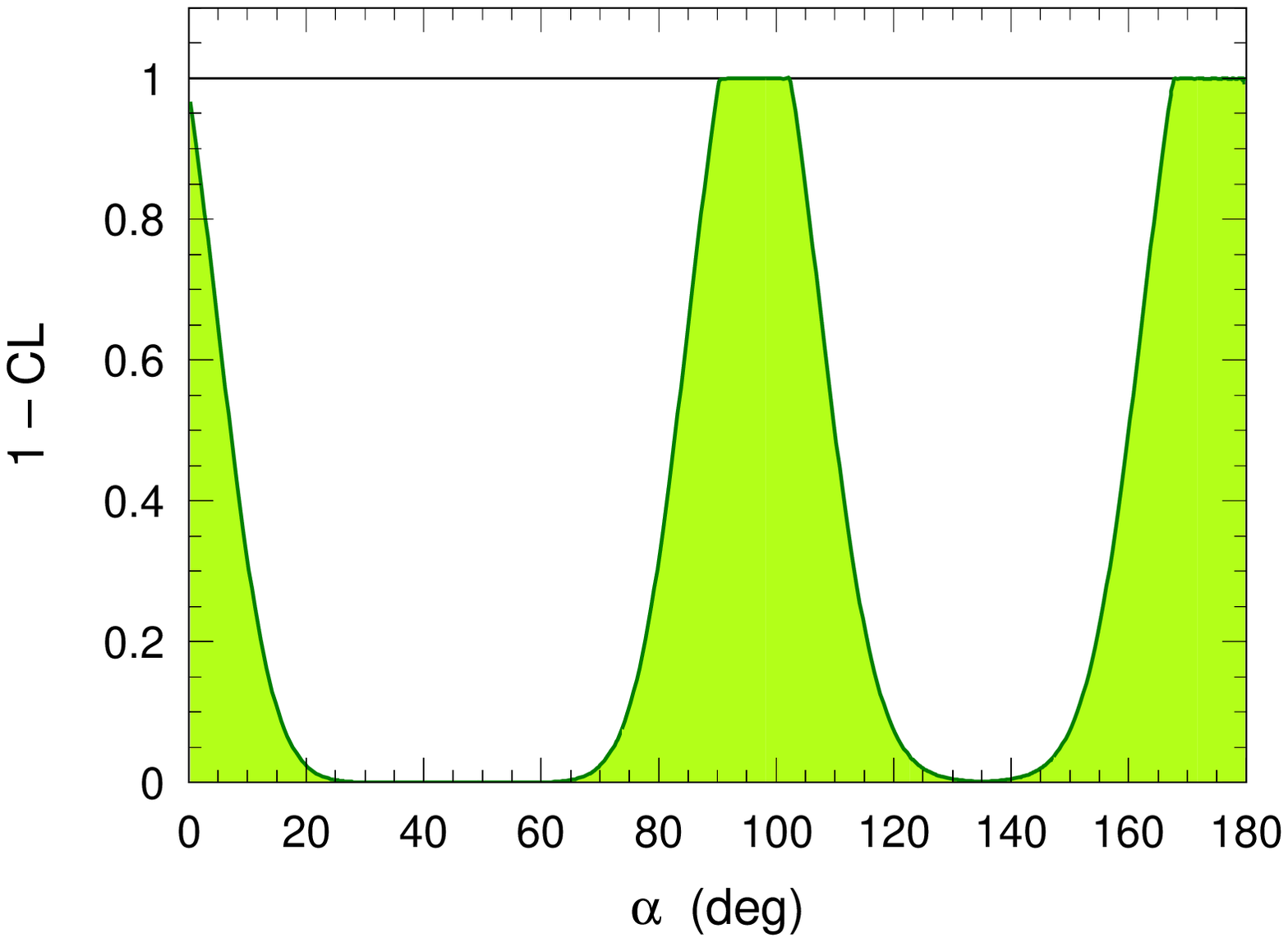} \\
\includegraphics[scale=0.43]{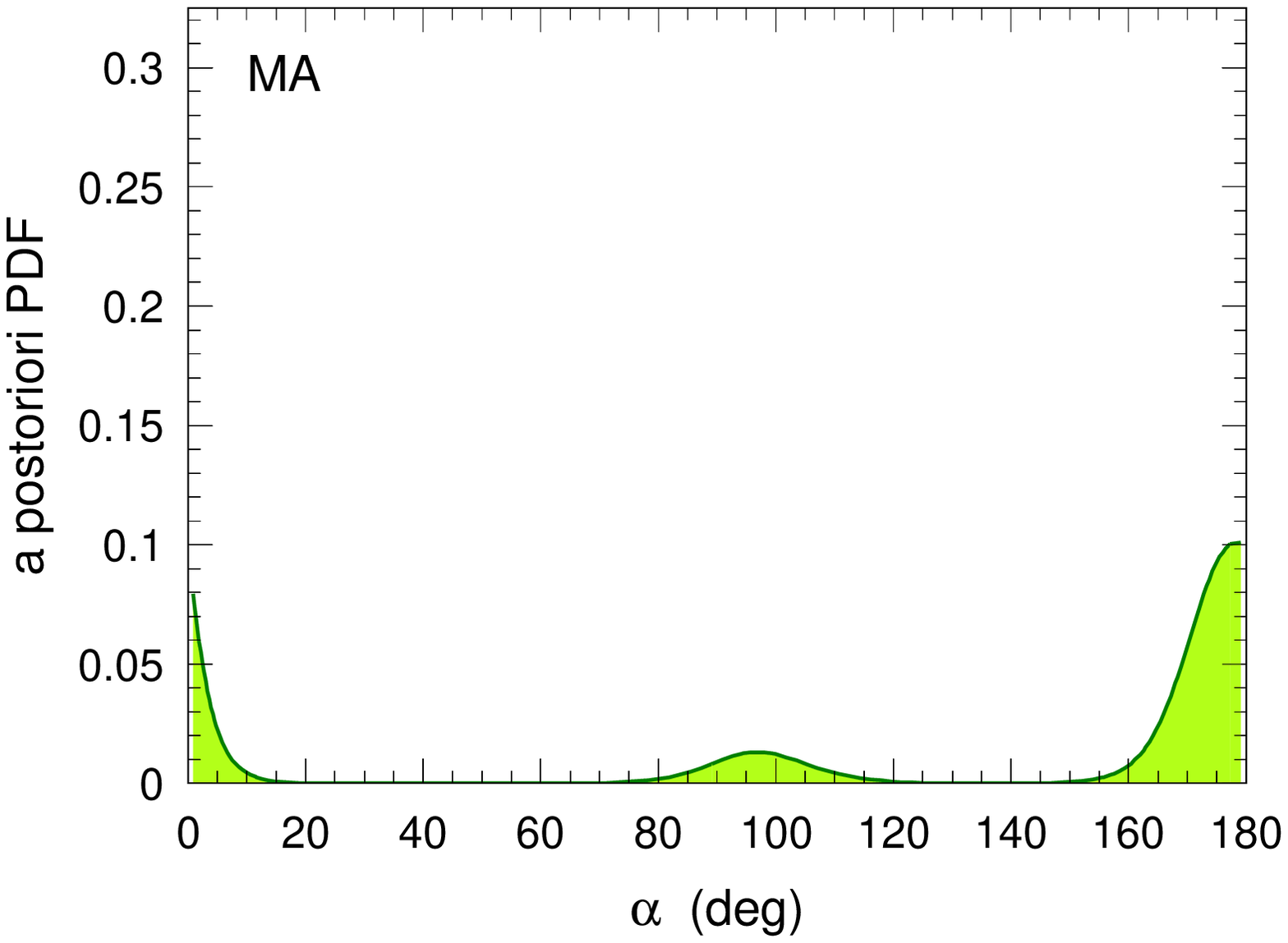}
\includegraphics[scale=0.43]{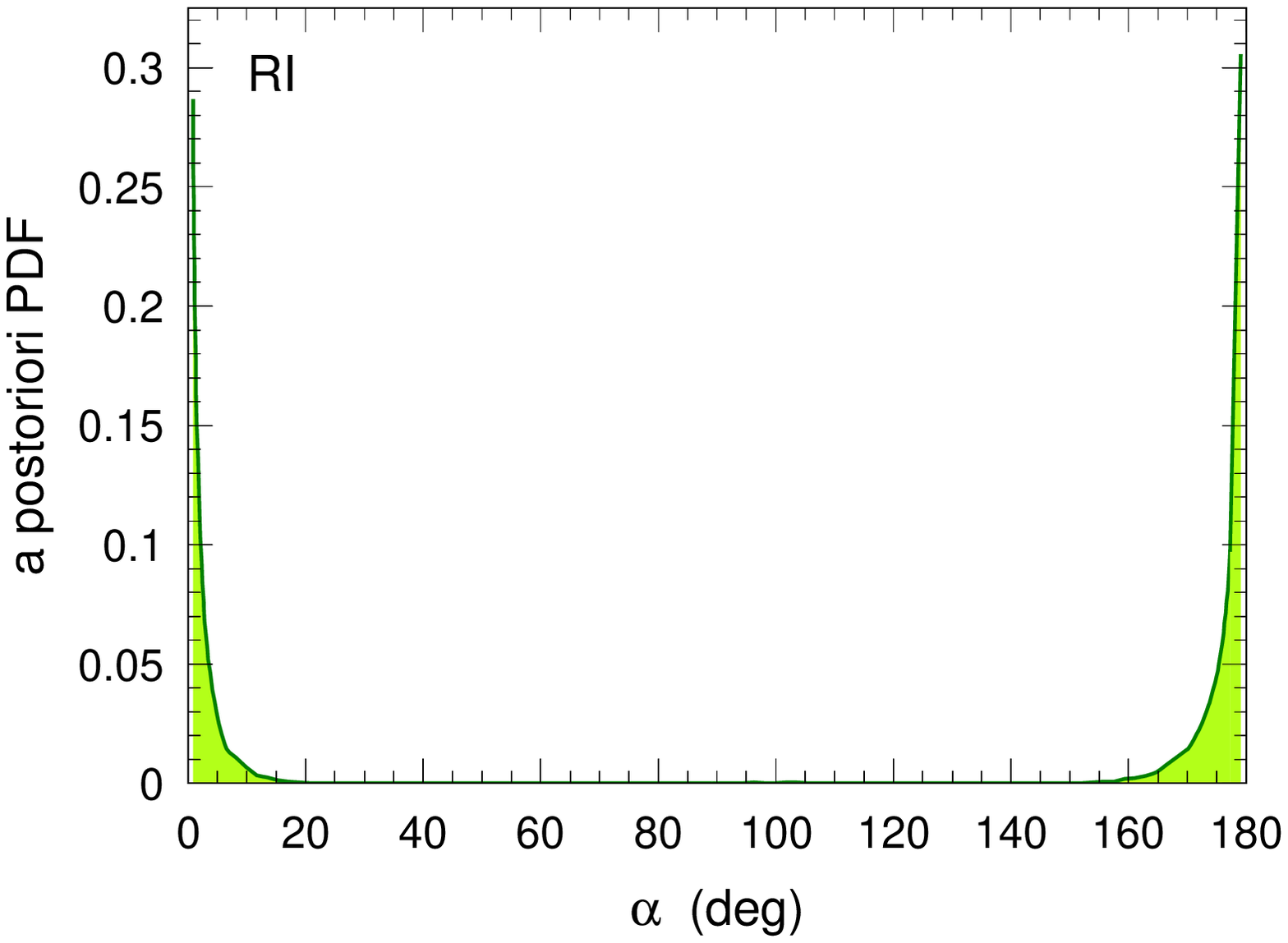} 
\includegraphics[scale=0.43]{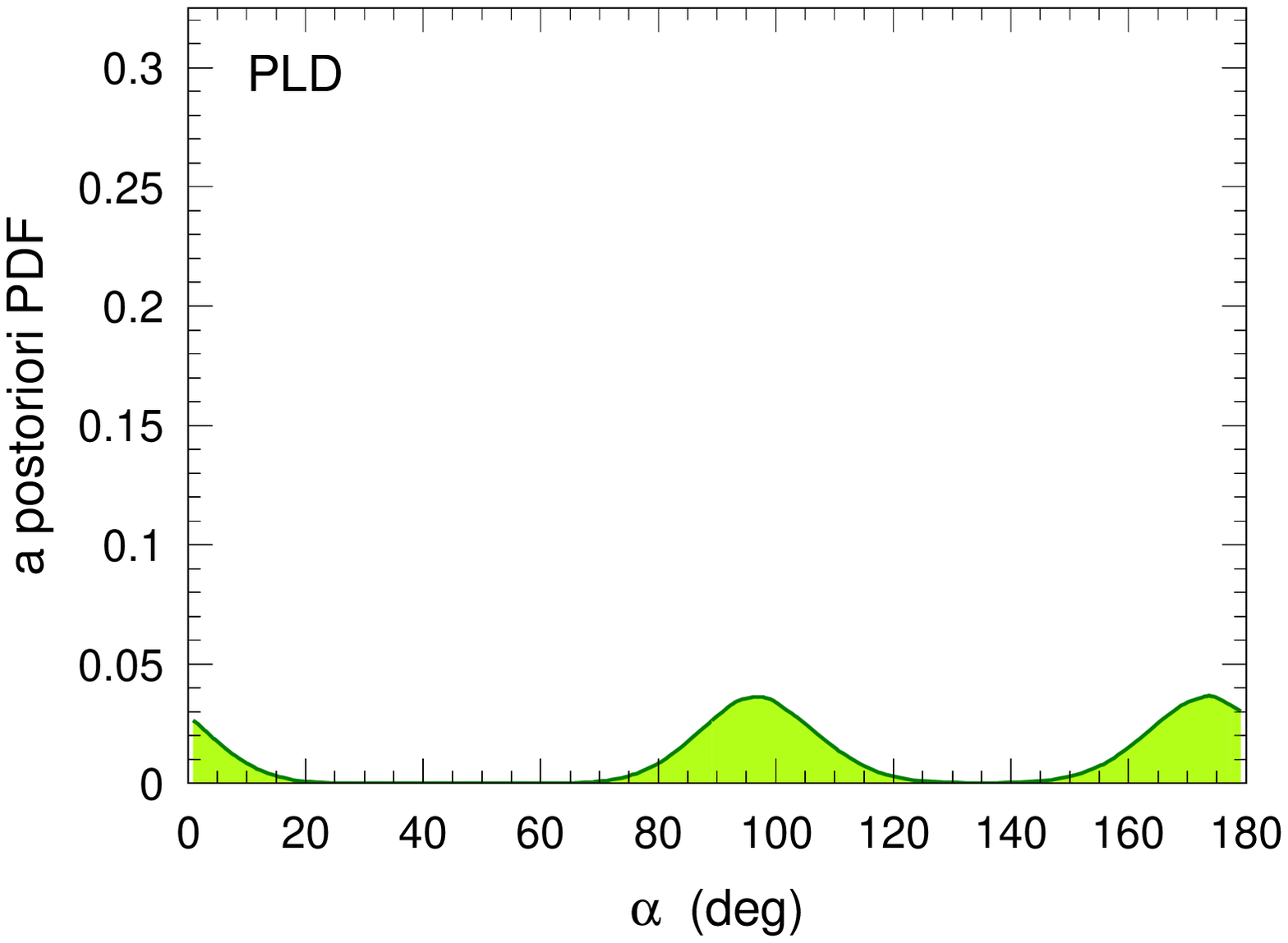}
\includegraphics[scale=0.43]{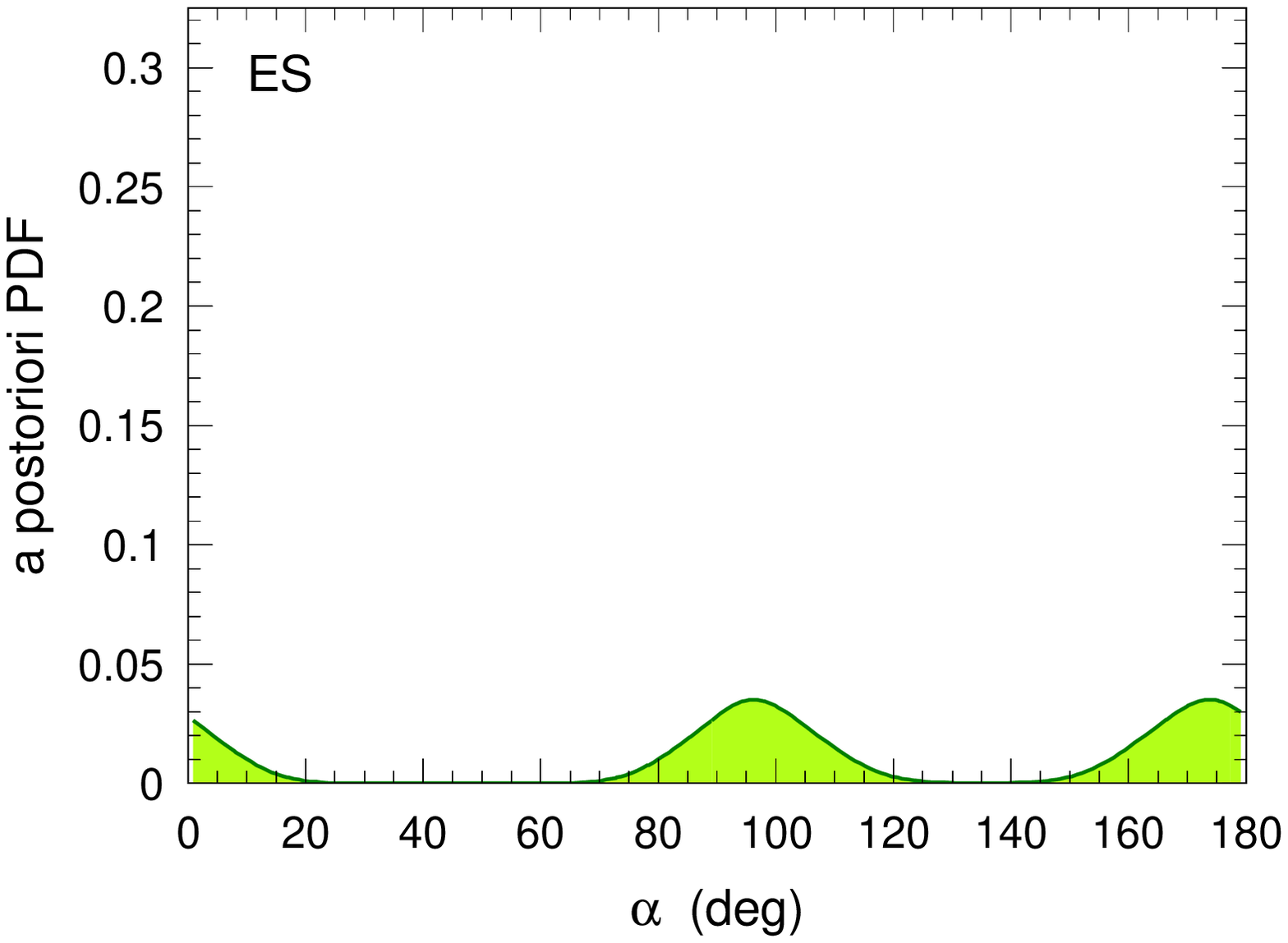}
\vspace{-0.3cm}
\caption{Results for the CKM phase $\alpha$ obtained with the different
         parameterizations from the $\B\to\rho\rho$ isospin analysis. 
         The upper plot shows the frequentist confidence level, which is 
         independent of the parameterization used. The remaining plots
         show the Bayesian a posteriori \pdfs for the parameterizations
         indicated by the labels.} 
\label{fig2}
\end{figure*}

The (longitudinally polarized) $B \to \rho \rho$ system is similar to
$B\to \pi\pi$ except that only an upper bound on the branching fraction 
to $\rho^0\rho^0$ is currently available. 
The current world average values for the observables are 
$\BR^{+-} = (25.1\pm3.7)\times 10^{-6}$, 
$\BR^{+0} = (19.1\pm3.5)\times 10^{-6}$, 
$\BR^{00} = (0.54\pm0.41)\times 10^{-6}$, 
$C^{+-} = -0.02\pm0.17$, 
$S^{+-} = -0.22\pm0.22$~\cite{rhorho} 
(the longitudinal polarizations are $f^{+-}=0.966\pm0.025$ and $f^{+0}=0.96\pm0.06$).
The ranges for the 
flat prior \pdfs in the various parameterizations are summarized in
Table~\ref{table2}. The resulting posterior \pdfs for $\alpha$ are
shown in Fig.~\ref{fig2}. Again the frequentist treatment (see  top plot
in Fig.~\ref{fig2}) is independent of the parameterization used. The constraints on $\alpha$ are lumped together under the 
two plateaus in the 1-CL curve.

In the PLD and \ES parameterizations, the range of
plausible values for the parameter $\alpha$ respect the symmetry of
the problem. As explained in Appendix~\ref{appalpha}, the experimental data are
still consistent with no \CP violation, so the posterior
\pdf for $\alpha= 0 [\pi]$ does not vanish. However, in the Standard Model
parameterization, the
prior breaks the symmetry of the problem and leads to a disaster in
the \RI parameterization (see discussion in Appendix~\ref{appb}). In
both cases, information contained in the data are convolved with additional
information provided by the priors: the posterior is dominated by the priors 
and not by the data.

%
%

\section{The \boldmath$\alpha \to 0 $ limit }\label{app:alimit}

In the next two sections, we present two 
additional parametrizations that describe finite \CP violation with $\alpha=0$ 
and finite values for the remaining parameters, and that are mathematically 
equivalent to the Standard Model parametrization except at $\alpha=0$.

\subsection{Mathematics}\label{appalpha}

The point $\alpha= 0\ [\pi]$  is a particular point of the theory for
physical reasons. In the Standard Model it corresponds to
vanishing \CP violation. The compatibility of the current $\B\to\pi\pi$ data with this
point is only marginal, because the \CP asymmetries are found to be 
different from zero~\cite{pipi,hfag}.

As described in Sections~\ref{sec:freq} and~\ref{sec:bayes}, the
behavior of the two statistical approaches around $\alpha= 0\ [\pi]$
shows striking differences. In the $\B\to\pi\pi$ case, the frequentist
method provides no significant exclusion confidence level for values of 
$\alpha$ close to zero (the fact that Fig.~\ref{fig1} (upper plot) 
depicts a continuous $1-\CL$ line through the point $\alpha=0$ is an 
artifact of the chosen binning of the $\alpha$ scan that does not hit 
this singular point). Setting $\alpha$ to zero
in the Standard Model parameterization leads to a confidence level of
the order of $10^{-8}$, which reflects the presence of \CP violation in
the data. The frequentist confidence level as a function of $\alpha$
is thus \textit{discontinuous}, as explained below and in 
Appendix~\ref{app:physics}.

On the other hand,
whatever the values of the observables, the \pdf as a function of
$\alpha$ is continuous at the origin in the Bayesian approach, and
exhibits a clear drop in the \MA parameterization. In the $\rho\rho$ case, where the experimental
data are still compatible with the absence of \CP violation, the
$\alpha= 0\ [\pi]$ value has a fairly good \pdf/CL in both Bayesian and
frequentist methods (with however the usual strong prior dependence for 
the Bayesian result).

The fact that vanishingly small, but non zero, values of the \CP phase
$\alpha$ can generate finite \CP violation is easily seen as
follows. Let us take the branching fraction $\cal{B}^{+-}$ and \CP asymmetry
$C^{+-}$ as an example: in the Standard Model parameterization, they
can be written as
\begin{eqnarray}
\cal{B}^{+-} &=& |T^{+-}|^2+2\cos\alpha\,\mathrm{Re}(T^{+-}P^*)+|P|^2\,, \\
C^{+-} &=& \frac{2\sin\alpha\, \mathrm{Im}(P/T^{+-})}
{1+2\cos\alpha\,\mathrm{Re}(P/T^{+-})+|P/T^{+-}|^2}\,.
\end{eqnarray}
The limit $\alpha\to 0$, $|T^{+-}|,\,|P|\to\infty$, and $P/T^{+-}\to -1$
is 
in general indeterminate as far as the observables are concerned. In particular any
finite value can be accommodated in this limit. The complete inspection
of the six observables shows that the combined limit\footnote
{
     Obviously, the divergences of some of the amplitudes
     is not an appealing physical result! However,
     in the present context of a set of six observables analyzed
     in the framework of SU(2) symmetry, nothing prevents its occurrence.
     In practice, one may add to the analysis new observables
     which could preclude these divergences. 
}
$\alpha\to 0$,
$|T^{+-}|,\,|T^{00}|,\,|P|\to\infty$ and $P/T^{+-},\,P/T^{00}\to
-1$ leads in general to finite branching fractions and \CP asymmetries. When
$\alpha$ is very small but non zero, this peculiar limit is fully
taken into account by the frequentist fit, which results in a finite
confidence level. In the Bayesian method, however, the  
priors mechanically suppress the above parameter configuration.

Finally we note that the limit $\alpha\to 0$ with finite observables
is obtained with finite values of the parameters
$\alpha_\mathrm{eff}$, $\mu$, $a$, $\bar a$ and $\Delta$ in the
PLD parameterization: this is due to the fact that
$\alpha_\mathrm{eff}$ can generate \CP violation even with $\alpha$
being strictly set to zero. The Standard Model and the PLD
parameterizations are equivalent except at the point $\alpha=0$, where 
the Jacobian corresponding to the change of variables is singular.

%
%
\subsection{Physics}\label{app:physics}

The parameterization in Eq.~(\ref{SMparam}) naively holds only within the
Standard Model. Let us assume arbitrary new physics contributions to
the $\Delta I=1/2$ channel; the most general parameterization can be
written as
\begin{eqnarray}\label{SMNPparam}
A^{+-} & =& e^{-i\alpha}T^{+-}+P +\MNP\,,
\nonumber\\
\sqrt{2} A^{00} & =&
e^{-i\alpha}T^{00}-P-\MNP\, ,\\
\sqrt{2} A^{+0} & =&
e^{-i\alpha}(T^{00}+T^{+-}) \,,\nonumber
\end{eqnarray}
where $\MNP$ is a complex  $\Delta I=1/2$ new physics amplitude.
For the $\Bzb$ decay the amplitudes above are \CP-transformed
in the following way: $\alpha\to-\alpha$, $\MNP\to\MNPb$.
Thus in general $\MNP\ne\MNPb$ and the new physics
contribution also violates \CP.

It is now clear that, if $\alpha\ne 0$ or $\pi$, Eq.~(\ref{SMNPparam}) can be
recasted into the Standard Model form (see Eq.~(\ref{SMparam}))~\cite{HQ}, with
$(T,P)\to(\tilde T,\tilde P)$ and
\begin{eqnarray}\label{repamInvariance}
\tilde T &=& T
+\frac{\MNPb-\MNP}{2i\sin\alpha} \,,\nonumber\\
 \tilde P &=& P
+\frac{e^{i\alpha}\MNP-e^{-i\alpha}\MNPb}{2i\sin\alpha}  \,.
\end{eqnarray}
In other words the Standard Model parameterization of the isospin
analysis is mathematically equivalent to a general Standard Model + new
physics parameterization, provided that the new contributions are purely
$\Delta I=1/2$. This equivalence is exact and holds for any value of the
parameters, except for $\alpha= 0\ [\pi]$.

This equivalence has an important physical consequence: it means that the
isospin analysis for the phase $\alpha$ bears no information on the
$\Delta I=1/2$ channel. In particular it may happen that \CP violation is
observed in specific asymmetries  such as $S^{+-}$, $C^{+-}$, $C^{00}$,
whereas $\alpha$ remains compatible with zero or $\pi$:
this configuration corresponds to a situation in which \CP violation is
generated by non-standard contributions to the $\Delta I=1/2$ channel.
Since the Standard Model parameterization already contains implicitly this
possibility, there is no need to add a specific new physics term to the
fit.

\begin{figure}[tb]
\includegraphics[width=0.47\textwidth]{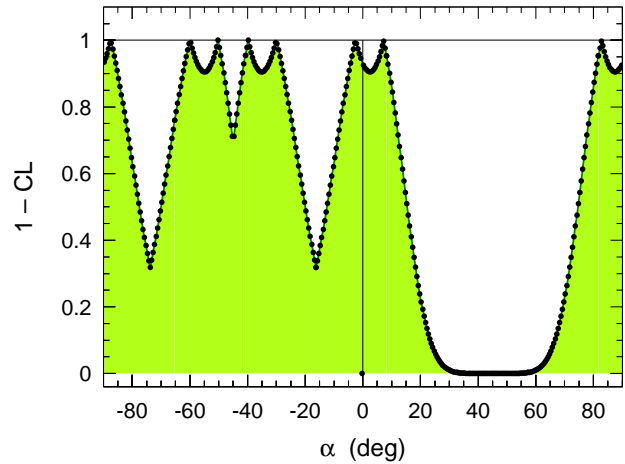}
\vspace{-0.6cm}
\caption{Fit results of the isospin analysis to the $B\to\pi\pi$ observables. Black dots:
Standard Model parameterization (see Eq.~(\ref{SMparam})); shaded curve: the
 same but allowing arbitrary new physics contributions to the $\Delta I=1/2$ channel (see Eq.~(\ref{repamInvariance})).}
\label{AlphaToZero}
\end{figure}
Figure~\ref{AlphaToZero} shows the result of the frequentist fit\footnote
{
	Note that this is a non trivial example where one has less 
	observables (six) than parameters in Eq.~(\ref{SMNPparam}) (ten).
} 
for both parameterizations from Eqs.~(\ref{SMparam}) and~(\ref{SMNPparam}). 
As expected from the above discussion, the two
curves are strictly identical except at the $\alpha=0\ [\pi]$ points where
the confidence level is low in the Standard Model parameterization due to the presence
of \CP violation in the data.

In contrast to the frequentist approach, the Bayesian treatment violates
the equivalence described in Eq.~(\ref{repamInvariance}), because choosing finite
priors in one parameterization automatically restricts the phase space 
in the other parameterization. The Bayesian posterior \pdf for $\alpha$ in the
parameterization in Eq.~(\ref{SMNPparam}) would yield a fairly good probability density at
$\alpha\sim 0$, since the amplitudes $\MNP$ and $\MNPb$ can
generate \CP violation. 

It may appear legitimate to perform a fit to the $B\to\pi\pi$
data within the Standard Model, not allowing for new physics neither in the
$\Delta I=3/2$ nor the $\Delta I=1/2$ transitions. From the above discussion 
it is not completely possible, unless amplitudes are known exactly within
the Standard Model. Still it is possible within the frequentist framework to restrict
the parameter space to ``reasonable'' ranges.
Regardless of the accuracy of these ranges, this would
not be the isospin analysis anymore since the isospin symmetry only
predicts Eq.~(\ref{SMparam}) without telling us anything about the order
of magnitude of the transition amplitudes.

\subsection{The \RI parameterization}\label{appb}

Mathematically, the fact that the \RI parameterization leads to
cataclysm in the Bayesian treatment can be understood as follows.
For a given set of model parameters $T^{+-}$, $P$, $T^{00}$ in Eq.~(\ref{SMparam}) one can always define the 
secondary parameters $\tau^{+-}$, $p$, $\tau^{00}$ such that:
\begin{eqnarray}\label{tauParam}
T^{+-} &=& \tau^{+-}/\sin\alpha\,,\nonumber\\
P &=& p-\tau^{+-}/\tan\alpha\,,\\
T^{00} &=& \tau^{00}-\tau^{+-}/\sin\alpha\,.\nonumber
\end{eqnarray}
The expressions of the three amplitudes in Eq.~(\ref{SMparam}) then take the form
\begin{eqnarray}\label{Tauparam}
A^{+-}&=& -i\tau^{+-} +p \,,\nonumber \\
\sqrt{2} A^{00}&=&  i\tau^{+-} +e^{-i\alpha}\tau^{00} -p \,, \\ 
\sqrt{2} A^{+0}&=&  e^{-i\alpha}\tau^{00} \,,\nonumber
\end{eqnarray}
where for the \CP-conjugate amplitudes one has to transform
$\alpha\to-\alpha$, $\tau^{+-}\to-\tau^{+-}$. It is now clear that $\tau^{+-} \ne 0$ can
generate finite \CP violation even if $\alpha=0$.

\begin{figure}[tb]
\centering
\includegraphics[width=0.47\textwidth]{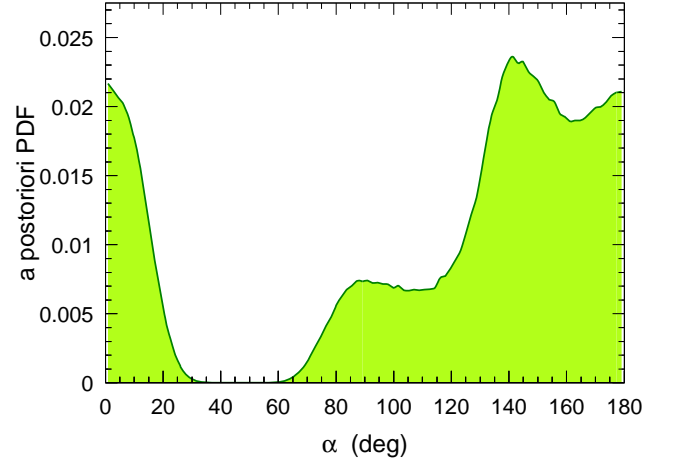}
\vspace{-0.6cm}
\caption
    {
      Posterior \pdf for $\alpha$ in the ``$\tau$'' parametrization (see Eq.~(\ref{Tauparam})) for $\B\to\pi\pi$ decays.
    }\label{fig:b5}
\end{figure}

The Jacobian of the transformation is
\begin{equation}
\left|
\frac{ d|\tau^{+-}| d\mathrm{Re}(p) d\mathrm{Im}(p) d\mathrm{Re}(\tau^{00}) d\mathrm{Im}(\tau^{00})}
{ d|T^{+-}| d\mathrm{Re}(P) d\mathrm{Im}(P)
 d\mathrm{Re}(T^{00}) d\mathrm{Im}(T^{00})}\right| = |\sin\alpha|
\end{equation}
and thus, in the \RI parameterization, the posterior for $\alpha$ can be written as:
\beq
P_{\mathrm{\RI}}(\alpha) \propto {1\over\mid\sin\alpha\mid} P_{\mathrm{\RIt}}(\alpha)
\eeq
with
\begin{eqnarray}
P_\mathrm{\RIt}(\alpha) &\propto& \int  e^{-{1\over 2}\chi^2\left[ |\tau^{+-}|,\mathrm{Re}(p),
\mathrm{Im}(p),\mathrm{Re}(\tau^{00}), \mathrm{Im}(\tau^{00}),\alpha \right] }\times \nonumber \\
& &\ \ d|\tau^{+-}| d\mathrm{Re}(p) d\mathrm{Im}(p) d\mathrm{Re}(\tau^{00}) d\mathrm{Im}(\tau^{00})\,. \nonumber
\end{eqnarray}
The $P_{\mathrm{\RIt}}$ function is the posterior \pdf for $\alpha$ in the ``$\RIt$'' parameterization. 
It is shown in Fig.~\ref{fig:b5} with the $B\to \pi \pi$ observables. One observes that $P_{\mathrm{\RIt}}$ is regular and finite 
for $\alpha=0$ and $\alpha=\pi$. As a result, 
it follows that in the \RI parameterization the posterior $P_{RI}(\alpha)$ is not only divergent at the origin (and for
$\alpha=\pi$) but is an improper \pdf: it cannot be normalized to unity. 
Applying the same treatment to the \MA parameterization exhibits no singularities as the one obtained in 
the \RI parameterization: the Jacobian $|d\RIt/d\mathrm{\MA}|$ behaves like $|\tau^{+-}|^2/\alpha$ at the origin which ensures 
that the posterior \pdf for $\alpha$ in the \MA parameterization vanishes at $\alpha=0^\circ$ ($\alpha=180^\circ$) 
in agreement with Fig.~\ref{fig1}.

\begin{figure}[tb]
\centering
\includegraphics[width=0.47\textwidth]{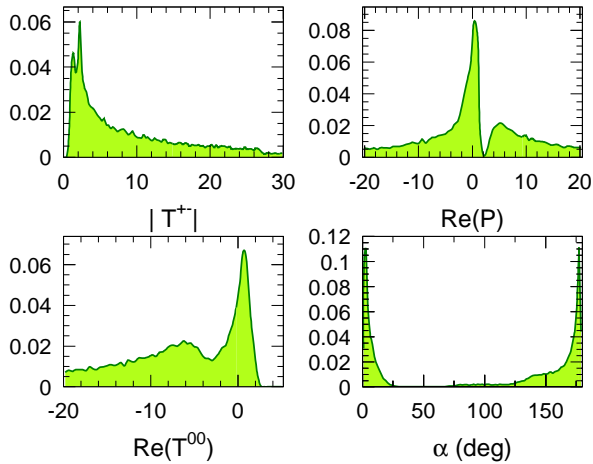}
\vspace{-0.6cm}
\caption
    {
      Posterior \pdfs in the \RI parameterization as obtained when widening the ranges of the priors 
      for $|T^{+-}|$, $\Re(P)$ and $\Re(T^{00})$ by a factor two compared to the ones used in Section~\ref{sec:param}.
      One observes that the posterior \pdfs for $|T^{+-}|$, $\Re(P)$ and $\Re(T^{00})$ do not vanish at the
      extremities of these extended ranges. In effect, the ranges of the prior must be extended
      to infinity, with the result that the posterior for $\alpha$ becomes an improper \pdf,
      with singularities at $\alpha=0^\circ$ and $\alpha=180^\circ$.
    }\label{fig:b6}
\end{figure}

As an illustration of this problem, the ranges of the priors for $|T^{+-}|$, $\Re(P)$ and $\Re(T^{00})$ in Eq.~(\ref{SMparam}) 
have been widened 
by a factor two compared to the ones used in Section~\ref{sec:param}. The corresponding posterior \pdfs are 
shown in Fig.~\ref{fig:b6}. One observes that the posterior \pdfs are truncated and 
that the posterior \pdf for $\alpha$ has its two peaks higher and shifted towards the endpoints ($0^\circ$ and $180^\circ$) 
with respect to the ones of Fig.~\ref{fig1}. It is worth noticing that the three ranges must be extended simultaneously 
to detect this effect.

%
%

\end{document}